\documentclass[a4paper,12pt]{article}

\usepackage{xurl}
\usepackage{tabularx}
\usepackage[pdftex]{graphicx}
\usepackage{epstopdf}
\usepackage{array}
\usepackage{longtable}
\usepackage{authblk}
\usepackage{fancyhdr}
\usepackage{titlesec}
\usepackage{hyperref}
\usepackage{caption}
\usepackage{amsmath}
\usepackage{tcolorbox}
\usepackage{times}
\usepackage{geometry}
\usepackage{dcolumn}
\usepackage{bm}
\usepackage{graphicx}
\usepackage{colortbl}
\usepackage{xcolor}
\usepackage{soul}
\usepackage{url}
\usepackage{tikz}
\usetikzlibrary{shapes, arrows, positioning}

\usepackage{pgfplots}
\pgfplotsset{compat=1.18}

\definecolor{pastelblue}{RGB}{174, 198, 207}
\definecolor{pastelgreen}{RGB}{119, 221, 119}
\definecolor{pastelyellow}{RGB}{255, 228, 130}
\definecolor{pastelbrown}{RGB}{166, 118, 91}
\definecolor{pastelpink}{RGB}{255, 182, 193}
\definecolor{pastelorange}{RGB}{255, 179, 71}
\definecolor{coral}{RGB}{255, 86, 83}
\definecolor{grey}{RGB}{233, 233, 226}
\definecolor{black}{RGB}{16, 17, 16}
\definecolor{tangerine}{RGB}{255, 128, 77}
\definecolor{mustard}{RGB}{255, 182, 73}
\definecolor{bordergrey}{RGB}{200, 200, 200}

\usepackage{tocloft}

\newcolumntype{M}[1]{>{\centering\arraybackslash}m{#1}}

\titleformat{\section}{\normalfont\Large\bfseries}{\thesection}{1em}{}
\titleformat{\subsection}{\normalfont\large\bfseries}{\thesubsection}{1em}{}

\hypersetup{
    colorlinks=true,
    linkcolor=blue,
    filecolor=magenta,      
    urlcolor=blue,
    pdftitle={Commodification of Compute},
    pdfpagemode=FullScreen,
}

\title{Commodification of Compute}
\author{Jesper Kristensen, David Wender, and Carl Anthony}
\affil{Symmetric Research}
\affil{\{jkristensen, dwender, canthony\}@symres.ai}

\date{June, 2024; v1.4}
\begin{document}

\maketitle

\begin{abstract}
The rapid advancements in artificial intelligence, big data analytics, and cloud computing have precipitated an unprecedented demand for computational resources. However, the current landscape of computational resource allocation is characterized by significant inefficiencies, including underutilization and price volatility. This paper addresses these challenges by introducing a novel global platform for the commodification of compute hours, termed the Global Compute Exchange (GCX\texttrademark) (Patent Pending). The GCX leverages blockchain technology and smart contracts to create a secure, transparent, and efficient marketplace for buying and selling computational power. The GCX is built in a layered fashion, comprising Market, App, Clearing, Risk Management, Exchange (Offchain), and Blockchain (Onchain) layers, each ensuring a robust and efficient operation. This platform aims to revolutionize the computational resource market by fostering a decentralized, efficient, and transparent ecosystem that ensures equitable access to computing power, stimulates innovation, and supports diverse user needs on a global scale. By transforming compute hours into a tradable commodity, the GCX seeks to optimize resource utilization, stabilize pricing, and democratize access to computational resources. This paper explores the technological infrastructure, market potential, and societal impact of the GCX, positioning it as a pioneering solution poised to drive the next wave of innovation in commodities and compute.
\end{abstract}

\newpage
\tableofcontents
\newpage



\section{Introduction}

\begin{figure}[t!]
    \centering
    \includegraphics[width=0.7\linewidth]{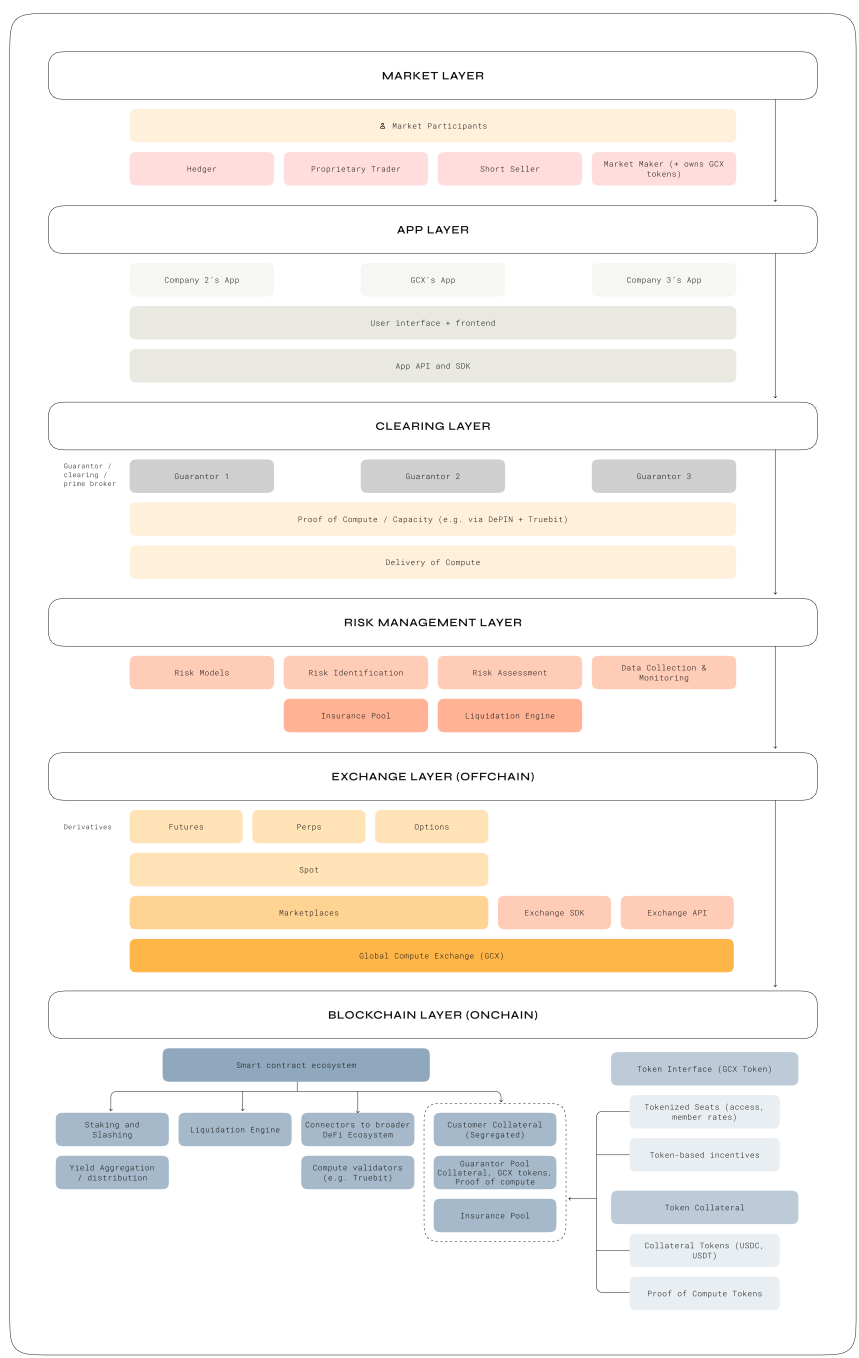}
    \caption[The Global Compute Exchange (GCX) Platform architecture.]{The Global Compute Exchange (GCX) Platform architecture is divided into several layers, each with specific functions. The Market Layer includes participants such as hedgers, traders, and market makers. The App Layer hosts various applications and user interfaces for interaction with the GCX. This also contains the GCX App and has an API and SDK for other apps to be built in this layer offering trading of compute to users. The Clearing Layer consists of guarantors providing proof of compute capacity and ensuring delivery. The Risk Management Layer features a risk engine with various components. The Exchange Layer (offchain) operates the core trading functions for compute resources. The Blockchain Layer (onchain) contains the smart contract ecosystem, including staking, liquidation engines, DeFi connectors, compute validators, customer collateral, and insurance pools. This multi-layered structure ensures a secure, efficient, and transparent market for trading compute resources.}
    \label{fig:gcx_diagram}
\end{figure}

In today’s rapidly evolving digital landscape, the demand for computational resources has escalated exponentially, driven by advancements in fields such as artificial intelligence, big data analytics, and complex scientific simulations. This surge has highlighted a significant challenge: the need for accessible and scalable compute power.
In response, we're building a Global Compute Exchange (GCX\texttrademark) (Patent Pending), see Fig.~(\ref{fig:gcx_diagram}), addressing this critical issue by creating a decentralized marketplace that not only democratizes access to these resources but also ensures efficient, scalable, and fair compute power distribution.

The demand for computational resources, particularly GPUs, has surged dramatically due to advancements in AI and machine learning, see Tab.~(\ref{tab:rpc_providers_summary}) and Fig.~(\ref{fig:demand}) \cite{sevilla2022compute}. This increased demand has resulted in a significant supply crunch, making computational power both expensive and difficult to access for many startups and independent developers. From a16z in \cite{Appenzeller2023}: ``There is no sign that the GPU shortage we have today will abate in the near future." Nvidia argues that the shortage of compute resources is not only due to a lack of GPUs themselves but rather their unavailability due to being tied up in contracts \cite{Constine2023}.

The ongoing compute shortages have compelled companies to innovate in order to maintain access to essential resources with firms seeking to reduce their GPU requirements. In 2023 alone, investors channeled hundreds of millions of dollars into startups focused on developing software that maximizes the efficiency of existing GPU resources. According to Chetan Kapoor from AWS \cite{awsdocs}, ``If there’s an ask from a particular customer that needs 1,000 GPUs tomorrow, that’s going to take some time for us to slot them in. But if they are flexible, we can work it out."
This flexibility and innovation are crucial as the industry navigates the challenges posed by limited GPU availability \cite{Dave2023}.

The rapid growth of ChatGPT \cite{ChatGPT} and the AI sector has significantly stressed the semiconductor industry and its supply chain. Companies that anticipated this demand reserved GPUs early, causing prices to double since 2020. Consequently, while spot instance pricing is available, vendors struggle to fulfill allocations, and acquiring clusters is nearly impossible without insider connections. This surge in software demand has outpaced the physical infrastructure's ability to produce hardware. The increasing complexity of chips, high-performance networking, and packaging has escalated prices, increased failure rates, and reduced yields.
Nvidia's \cite{nvidia2024} Robert Ober notes that despite cloud leaders' requests to boost production tenfold, the intricate nature of these systems prevents such rapid scaling. Improving maximum performance through optimized ``model flop utilization" and securely intermixing users to ensure constant hardware operation is essential as we work towards scaling up compute manufacturing \cite{Constine2023}.

The concept that ``compute is the new oil" encapsulates this paradigm shift, highlighting the critical role that computational resources play in driving technological innovation and economic growth \cite{timofeev2024case, jindal2024compute}.

\begin{figure}[t!]
    \centering
    \includegraphics[width=\linewidth]{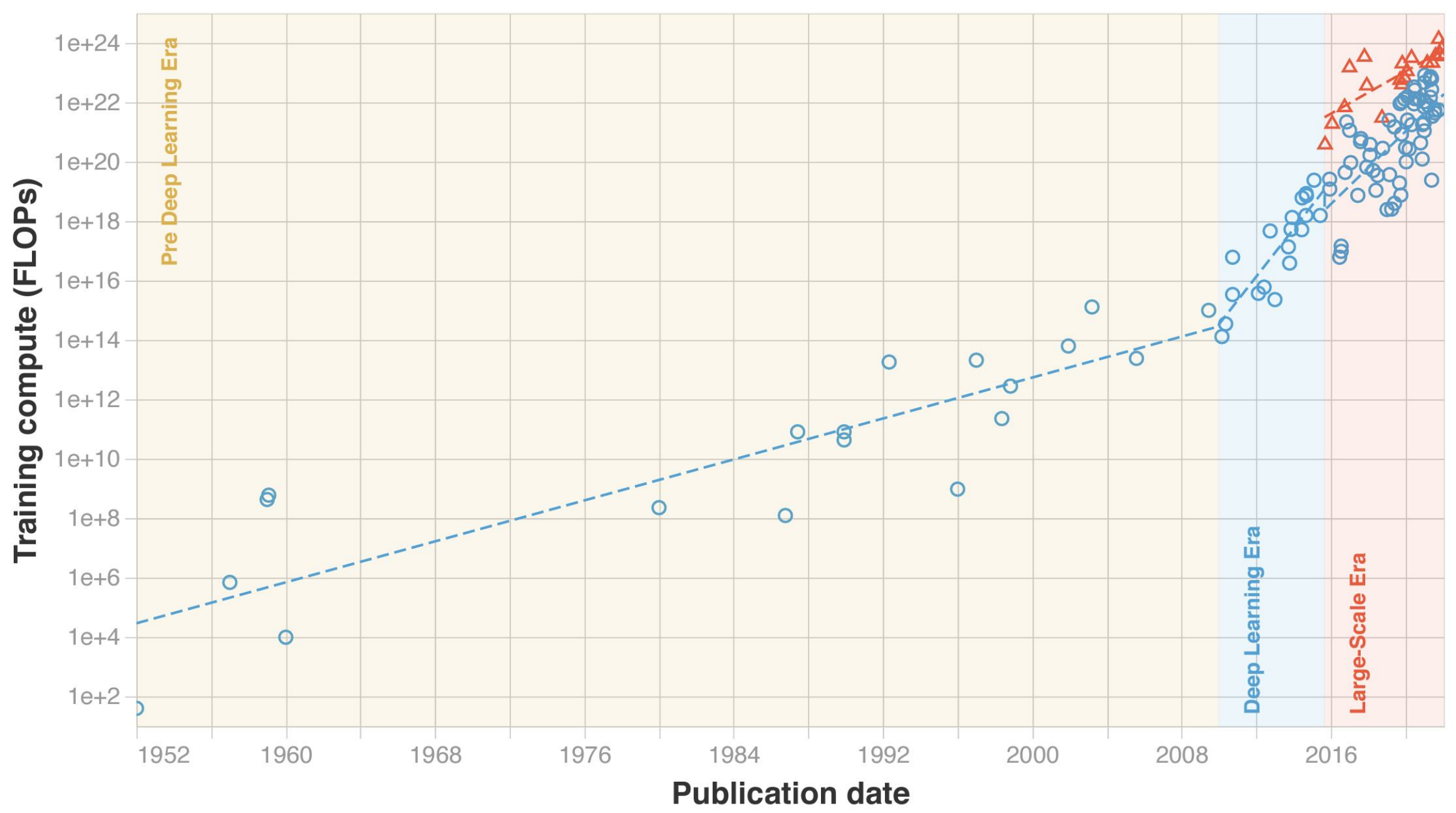}
    \vspace{0.5\baselineskip}
    \caption{Trends in FLOPs for 121 milestone ML models between 1952 and 2022. The graph illustrates the growth in computational requirements (FLOPs) over time, highlighting three distinct eras. A noticeable change in slope occurs around 2010, corresponding to the advent of Deep Learning. Additionally, a new large-scale trend emerges in late 2015, indicating further advancements in ML model complexity. Replicated from \cite{sevilla2022compute}.}
    \label{fig:demand}
\end{figure}

\begin{figure}[t!]
\vspace{2\baselineskip}
\captionsetup{name=Table}
\setcounter{figure}{0}
\captionsetup{list=off} 
\centering
\includegraphics[width=\linewidth]{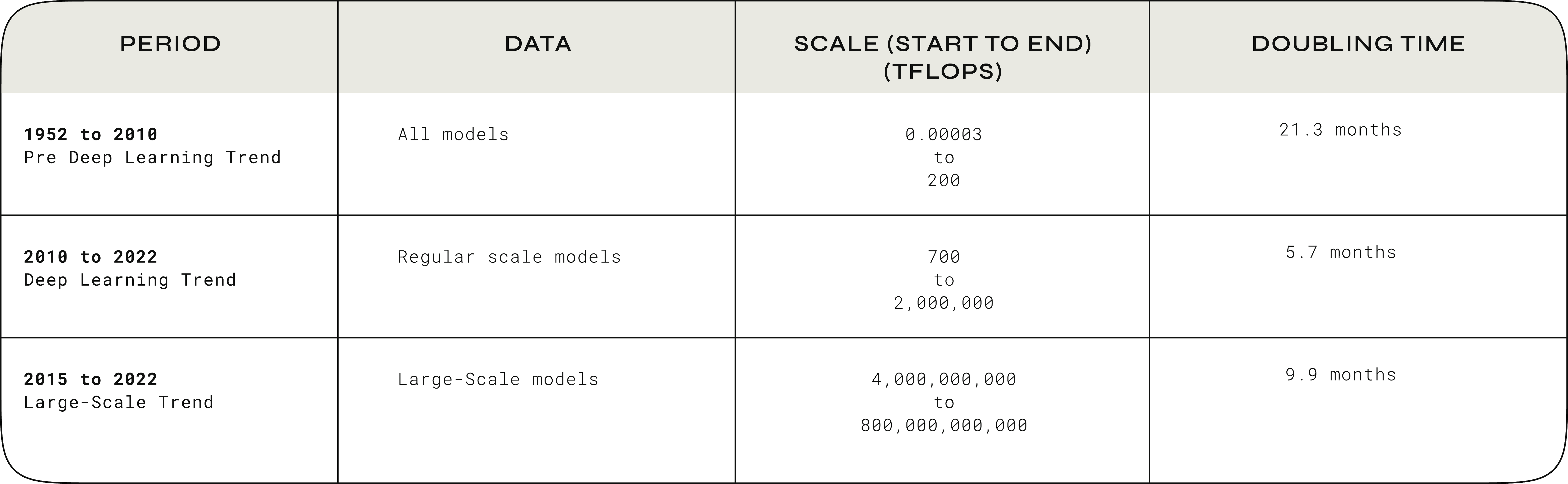}
\caption{Compute trends across three eras of machine learning. The table presents the periods, the data set size, the scale of compute from start to end, and the corresponding doubling times. In the Pre Deep Learning \cite{lecun2015deep} era (1952 to 2010), all models exhibit a scale increase from 0.00003 to 200 TFLOPs with a doubling time of 21.3 months. The Deep Learning era (2010 to 2022) saw regular-scale models increase from 700 to 2,000,000 TFLOPs with a significantly faster doubling time of 5.7 months. Finally, the Large-Scale era (September 2015 to 2022) involved large-scale models scaling from 4,000,000,000 to 800,000,000,000 TFLOPs with a doubling time of 9.9 months. From \cite{sevilla2022compute}.}
\label{tab:rpc_providers_summary}
\end{figure}

As AI, and general cloud compute, adoption continues to accelerate, the need for scalable and efficient compute resource management becomes ever more critical.
What is more, research indicates that the computational requirements for leading-edge AI systems, for example, have been doubling approximately every three to four months \cite{openai2023,Thompson2020}, see again Tab.~(\ref{tab:rpc_providers_summary}). Such growth rates substantially outpace Moore's Law \cite{schaller1997moore}, emphasizing the urgent need for innovative solutions in resource provisioning and management \cite{brookings2018,guo2022}.

Current projections indicate that the demand for compute power will continue to grow exponentially, driven by the increasing complexity of AI models and the widespread adoption of data-intensive applications \cite{guo2022edge}. According to Shoal Research \cite{shoalresearch}, the demand for computational resources, particularly GPUs, has surged, causing a supply crunch and making it expensive and difficult for startups and independent developers to access necessary resources \cite{timofeev2024case,brown2022dynamics}. This surge underscores the critical need for innovative solutions to efficiently manage and distribute these resources, ensuring that technological innovation is not hindered by resource constraints.

We introduce a marketplace that leverages blockchain technology to create a transparent, secure, and efficient platform for trading compute resources \cite{xu2017}. By using a token-based economy, we incentivize the proper allocation and utilization of resources, with mechanisms for penalties and rewards that ensure compliance and performance \cite{ottina2023}. As noted by Ren et al. \cite{Ren2009}, such approaches are crucial for managing billion- and trillion-scale model training sessions that require substantial computational efforts \cite{underwood2018}.
The Blockchain Layer also enables the implementation of innovative Decentralized Finance (DeFi) trading technologies for compute resources, such as perpetual futures (perps) \cite{he2022fundamentals} and options.

Advanced monitoring and predictive analytics are integrated to ensure proper risk management and that compute delivery is both reliable and efficient \cite{biewald2020}. The platform's capability to provide real-time analytics helps preempt potential failures, ensuring uninterrupted service delivery \cite{kim2016}. This reliability is essential not only for maintaining service quality but also for building trust within the marketplace.

\begin{figure}[t!]
    \centering
    \includegraphics[width=\linewidth]{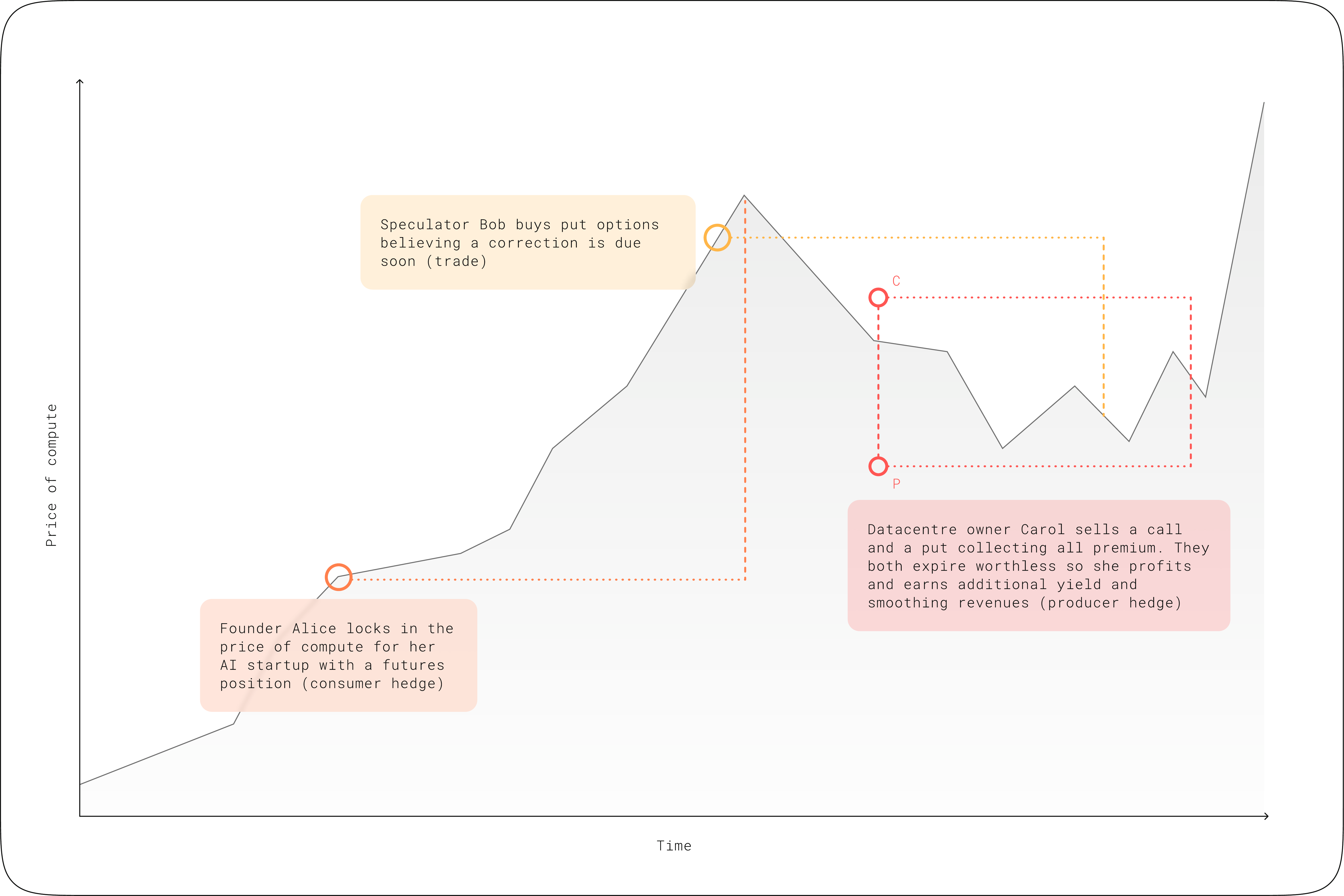}
    \caption[Trading strategy examples used by various market participants in the compute market]{This figure illustrates different strategies used by various market participants in the compute market. Founder Alice locks in the price of compute for her AI startup using a futures position, effectively hedging against future price increases (consumer hedge). Speculator Bob buys put options, anticipating a market correction to secure a price floor and minimize potential losses (trade). Datacenter Owner Carol sells both call and put options, collecting premiums. When these options expire worthless, she profits from the premiums, generating additional yield and smoothing her revenue stream (producer hedge). The figure demonstrates how these strategies can benefit different actors in the market by managing risk and optimizing financial outcomes \cite{hull2018options}.}
    \label{fig:hedging_scenarios}
\end{figure}

The GCX provides a versatile platform that caters to various stakeholders in the compute market, enabling them to hedge against price volatility, secure future prices, and generate additional yield. As illustrated in Fig.~(\ref{fig:hedging_scenarios}), the GCX allows different actors to leverage its functionalities to their advantage. Alice, an AI startup founder, uses futures contracts to lock in the price of compute, safeguarding against future price increases \cite{kolb2016futures}.
We already see glimpses of this happening albeit in a fragmented and inefficient way compared to having a centralized marketplace: In 2023, for example, Resemble turned to FluidStack, a tiny provider that welcomes one-week or one-month GPU reservations, and has recently joined San Francisco Compute Group, a consortium of startups jointly committing to buy and split GPU capacity \cite{Dave2023}.
Bob, anticipating a market correction, buys put options to secure a price floor, thereby minimizing potential losses \cite{hull2018options}. Carol, a data center operator, sells both call and put options to collect premiums, generating additional yield. When these options expire worthless, she profits from the premiums, smoothing her revenue stream (producer hedge) \cite{passarelli2012trading}. By utilizing the GCX, these actors can effectively manage their risks and optimize their financial strategies in the dynamic compute market. 

Unlike commodities such as oil and coffee, compute has not yet become a tradable commodity in established markets. The absence of compute commodity markets may be attributed to several factors, including the rapid evolution of technology, the complexity and diversity of compute resources, and the lack of standardized units for trading compute.
In all cases, notice that the GCX enables various expressions of market sentiment, allowing participants to profit and protect against the evolution of compute prices, whether they anticipate an increase, decrease, or sideways movement. This flexibility ensures that the GCX can benefit a diverse range of market participants, each with their unique perspectives and strategies.

\textbf{What should the price of compute be?} This is not known. The price of compute can vary widely based on factors such as demand, supply, technological advancements, and market conditions. Unlike traditional commodities, the compute market is influenced by a unique set of variables, making it challenging to pinpoint a stable price. The GCX aims to bring transparency and stability to this nascent market, enabling participants to better understand and manage the cost of compute resources.
We're at the forefront of addressing the critical challenge of compute power scarcity and uneven distribution. By creating a marketplace that makes compute resources accessible and efficiently managed, we pave the way for a future where computational power is no longer a barrier to innovation but a widely available asset that drives global technological advancement \cite{dean2020}.

The on-demand accessibility to cutting-edge infrastructure, such as GPU clusters, FPGAs \cite{maxfield2004design}, and quantum computing \cite{Nielsen_Chuang_2010} from partners like Cambridge Compute Co. \cite{cambridge_compute_c3} into the GCX platform, combined with increased budget flexibility from cost savings, empowers organizations to accelerate ideas that were previously unattainable. Data teams can conduct experiments more rapidly and iteratively across diverse hardware types and geographic locations, leveraging cloud interoperability rather than being confined to local limitations.

Engineers can easily prototype by utilizing global capacity on demand, prioritizing innovation velocity over perfection. Startups, in particular, benefit immensely from this model, as it allows them to share excess resources inexpensively, avoiding significant capital investments while still validating product-market fit \cite{democratizing}.

In essence, by addressing the dual constraints of capacity and cost that hinder cloud consumption at scale, Compute as a Commodity (CaaC) introduces new levels of freedom that enhance creativity, productivity, and outcomes. By aggregating fragmented infrastructure across isolated domains, CaaC achieves collective efficiency and democratizes access, thereby removing innovation bottlenecks.

In the following chapters, we will cover into the mechanics of standardizing compute units, grading compute resources, creating compute pools, ensuring delivery, managing risks, and building the trading platform, the GCX, that underpins this new marketplace.
\section{Evolution of Compute: From Mainframes to Compute as a Commodity}

The shift from capital-intensive on-premises infrastructure to operational expenditure models enabled by cloud computing has revolutionized access to compute resources \cite{ceruzzi1998moder,techtarget,ince2011computer}. Cloud computing, pioneered by companies like Amazon Web Services, has allowed businesses to rent capacity and scale rapidly without significant upfront investments \cite{cloudpwr2024,cloudflare}. This trend is paving the way for CaaC, which aims to make computing power as accessible and affordable as utilities like electricity and water \cite{democratizing}.

However, compute power has not truly been commoditized yet—this is where we step in to pioneer the next phase of this evolution.
By further building on the abundance of underutilized capacity across clouds and data centers, Compute as a Service (CaaS) takes the cloud consumption model to the next level for delivering even better economics and flexibility.
The coming CaaC revolution promises to further the cloud's democratization mission and make dynamically scalable computing an inexpensive commodity for fueling innovation of all scales.

The concept of cloud computing has been instrumental in enabling the CaaC model to emerge. By providing on-demand access to shared compute resources over the internet, cloud computing realizes the vision of utility computing—it allows organizations to consume fundamental computing services without massive infrastructure investments.
The origins of cloud computing can be traced back to the 1950s when large-scale mainframes were starting to gain adoption for critical enterprise workloads. Given their exorbitant costs, efficient utilization through sharing of compute time was necessary, leading to the development of optimization techniques like workload consolidation, virtualization, shared storage, and auto-scaling, which came to define the core tenets of cloud computing decades later.

By the 1990s, when data centers had proliferated for housing business applications, capacity utilization remained poor due to fragmented systems and intermittent traffic spikes. The internet boom in the 2000s also caused unprecedented growth in web-scale infrastructures built around horizontal scaling. Scholars began exploring models for ``computing utilities" much like electricity and water utilities where users could access computing functionality without regard for the underlying delivery.
The launch of Amazon Web Services’ (AWS) S3 storage and EC2 compute offerings in 2006 is considered by many as the symbolic start of the modern cloud computing era, which delivered on this promise of utility services \cite{awshistory}.
It allowed small teams and startups to bypass investing in servers and data centers by renting Amazon’s spare capacity, which had been built out for the retail giant’s peak usage. By exposing their large-scale systems through web services and pioneering the prepaid cloud pricing model, AWS led a paradigm shift that adopted economies of scale \cite{erl2013cloud}.

Over the next decade, many enterprises migrated select workloads into the AWS public cloud, attracted by flexibility, resilience, and usage-based billing instead of high fixed costs. The ability to get started with no upfront commitment removed barriers to test new ideas and go-to-market quicker.
Renting compute removed longer-term risks as well. Google, Microsoft, and others followed AWS’s footsteps by launching their own IaaS and PaaS products to woo customers. A multi-cloud world began taking shape. Today, over 90\% of enterprises have a multi-cloud strategy combining on-premises and off-premise environments.
Cloud has transitioned from experimentation to a core pillar of enterprise computing.

While significant advancements have been made, compute power has not yet been commodified.
With the GCX, we're at the forefront of this transformation, leveraging the abundance of underutilized capacity across clouds and data centers to take the cloud consumption model to the next level.
The coming CaaC revolution promises to make dynamically scalable computing an inexpensive commodity, fueling innovation of all scales. We are pioneering this next step, aiming to further the cloud’s democratization mission and make high-performance computing accessible for everyone.

\section{Commodification of Compute}

Compute commodification refers to the process of transforming computational power into a standardized, tradable asset \cite{commodification_wiki}.
Commoditization, on the other hand, refers to the process where goods become undifferentiated and subject to increased competition \cite{commoditization_wiki}. In this way, we set out to commodify compute to enable its commoditization.
Compute commodification involves defining discrete units of compute resources that can be bought, sold, and traded on a marketplace, similar to traditional commodities like electricity, oil, or metals \cite{timofeev2024case, jindal2024compute}.
The commodification of compute resources ensures efficient resource allocation and fosters competitive pricing. By creating a standardized measure of compute power, it becomes possible to commodify and trade these resources efficiently and transparently. Transparent and efficient marketplaces for computational resources can significantly enhance resource utilization and market dynamics \cite{smith2021democratizing, xu2017efficient}.

The core value proposition of CaaC boils down to ready availability of affordable compute capacity for organizations on tap, much like any utility service for modern life such as electricity or piped gas. Unlike traditional cloud providers such as AWS, which require significant setup and management, CaaC presents compute resources as an instantly usable operational expenditure. This model dynamically scales to workload needs without extensive configuration, reducing both the technical expertise required and the administrative burden on users.
Computing transitions from an owned asset requiring upfront capital allocation and skilled labor, to an instantly usable operational expenditure that auto-scales to workload needs dynamically \cite{democratizing}.

The benefits fundamentally stem from separating demand for computing innovation versus supply of
computing infrastructure. Enterprises are hence able to convert CapEx investments to more strategic OpEx spends \cite{investopedia2024capexopex} that help fuel differentiation for their digital capabilities instead of competing merely at host
hardware or data center levels with shrinking marginal utility \cite{democratizing}.

The commodification of compute resources offers several significant advantages.

Firstly, it enhances market liquidity. By standardizing compute units, the market becomes more liquid, with a greater number of buyers and sellers able to participate. This increased liquidity makes it easier to trade compute resources \cite{gibbons2020, balatsky2015}. Furthermore, blockchain technology and smart contracts enhance transparency and trust in transactions, providing a secure and reliable platform for trading compute resources \cite{Thompson2020, Schär2022}.

Secondly, the commodification of compute resources lowers transaction costs and enables market participants to express a more comprehensive outlook on the price of compute. This predictability and reduction in transaction costs benefit consumers and attract more providers to the market. By offering tools that cater to a diverse set of risk profiles, the GCX ensures that the market is accessible to a wide range of participants, from large-scale data centers to smaller, niche providers. Our goal is not to disrupt existing markets but to expand the overall pie, creating opportunities for all participants. The increased predictability and efficiency of the GCX platform support investment and growth in the compute market, making it an attractive space for innovation and expansion.

By activating dormant resources, the compute market could significantly expand. There is substantial underutilized computational power in personal devices like smartphones and PCs, which could be leveraged through financial incentives. Similarly, many companies have in-house compute resources that are often idle during non-peak hours, representing another untapped resource. Additionally, the commodification model could lower barriers for smaller providers, encouraging the emergence of startups and niche providers. This would foster a more competitive and diverse market, attracting investment and driving innovation.

To estimate the potential market size, consider the following contributions:
\begin{itemize}
\item \textbf{Personal Devices}: With an estimated 6.84 billion smartphones globally, contributing 1 GFLOP for 2 hours daily, the total annual revenue could reach \$499.32 billion based on current rates for cloud services.
\item \textbf{Personal Computers}: With an estimated 2 billion personal computers globally, contributing 10 GFLOP for 2 hours daily, the total annual revenue could reach \$1.46 trillion.
\item \textbf{In-House Company Resources}: If multiple corporations collectively contribute 1 exaFLOP of their compute capacity continuously, the annual revenue could be \$87.6 billion.

\item \textbf{Emerging Smaller Providers}: Assuming 10,000 new providers each contribute 100 TFLOPs, the annual revenue could also be \$87.6 billion.
\end{itemize}

Summing these contributions, the potential market size could reach more than \$2 trillion annually. This substantial economic potential underscores another significant benefit of commodifying compute resources.

Thirdly, commodification also provides flexibility and efficiency. Users can purchase compute power as needed, allowing for flexible scaling of resources based on demand. This dynamic allocation helps optimize resource usage and reduce wastage \cite{xu2017efficient, smith2018resource}.
It is estimated that about 76\% of non-production cloud resources (such as those used for development, staging, testing, and QA) remain idle during non-working hours. This translates to a substantial amount of compute resources being paid for without being actively used. Also, many instances are provisioned with more capacity than necessary. For example, around 40\% of cloud instances are sized larger than needed, leading to unnecessary expenses. Downsizing these instances could save significant costs \cite{chapel2024cloud,stagnitto2024cloud}.
Access to a marketplace for compute resources means that businesses and individuals can acquire the exact amount of compute power they need, when they need it, without long-term commitments \cite{dongarra2003high, dongarra2013high}.

Lastly, commodification promotes accessibility and inclusivity. Compute commodification democratizes access to high-performance computing, allowing smaller businesses and individuals to leverage powerful compute resources that were previously accessible only to large corporations \cite{smith2021democratizing, green2020democratization}. Easier access to compute power can spur innovation and development, enabling new applications and technologies that rely on intensive computational resources \cite{hinton2023future, openai2023}.
Quoting from \cite{Dave2023}: ``The startup ecosystem is trying to get together and try to figure out `How do we battle, how do we fight for compute?' Otherwise, it would be a really unfair game. Prices are just too high."

The substantial investments in AI alone, with over \$28 billion raised by private AI companies since 2020, underscore the critical need for efficient compute resource management systems. This economic potential drives the demand for commodifying compute resources to support AI growth. The world is going to go to 80, 90\% of all cycles, are going to be AI cycles running on AI processors programmed by data \cite{tenstorrent2024}.
The GCX is a marketplace where compute resources are as easily tradable as traditional commodities. We are building a platform that standardizes compute units, ensures reliable delivery through standardized contracts, and fosters a vibrant ecosystem of buyers and sellers. 

And standardization of compute will allow GCX to create a robust derivatives market which will allow for better price transparency and risk management for all market participants.
By commodifying compute power, we aim to unlock new opportunities for innovation, efficiency, and growth across various industries.

\section{Compute as the New Oil}

The demand for computational resources, especially GPUs, has surged dramatically due to advancements in AI and machine learning.
What is more, the compute resource market faces significant challenges, including resource allocation inefficiencies, data privacy and security issues, and competitive barriers for small and medium-sized enterprises.
The requirements for AI models have grown exponentially, with compute needs increasing by 70 times from GPT-3 to GPT-4. This highlights the urgent need for scalable and efficient compute resource management solutions to meet the growing demands of AI technologies.
This increased demand has resulted in a significant supply crunch, making computational power both expensive and difficult to access for many startups and independent developers. The concept that ``compute is the new oil" encapsulates this paradigm shift, highlighting the critical role that computational resources play in driving technological innovation and economic growth \cite{timofeev2024case, jindal2024compute}.

Decentralized Physical Infrastructure Networks (DePINs) have emerged as a solution to the challenges of accessing computational resources \cite{fetch2024depins}. By creating decentralized marketplaces, DePINs enable individuals and organizations worldwide to offer their idle compute power. This approach not only democratizes access to computational resources but also drives down costs through increased competition and resource utilization efficiency \cite{timofeev2024case}.

Despite their potential, compute DePINs face several economic and technical challenges. Competing with and working alongside centralized providers requires overcoming issues related to scalability, reliability, and cost-effectiveness. However, the integration of blockchain technology and smart contracts within these networks can help mitigate these challenges by providing secure, transparent, and efficient platforms for trading compute resources  \cite{mit2024breakthrough, nature2024ai}.

The imbalance between the supply and demand for compute resources presents a significant market opportunity. By leveraging decentralized solutions, platforms like the GCX can bridge this gap, making computational power more accessible and fostering a more inclusive environment for AI development. This democratization of compute power is crucial for enabling innovation across various sectors, from startups to educational institutions and research organizations \cite{timofeev2024case, mit2024cs}.

Building the GCX is vital for several reasons. Firstly, it addresses the growing demand for computational resources by creating a decentralized marketplace that can efficiently match supply with demand. This is essential for ensuring that the burgeoning field of AI and machine learning continues to advance without being hampered by resource constraints. Secondly, the GCX provides a platform that democratizes access to high-performance computing, allowing a broader range of participants, including startups, researchers, and institutions from economically disadvantaged regions, to access the computational power they need to innovate and grow \cite{mit2024cs, green2020democratization}.

Furthermore, the GCX can enhance the availability and affordability of computational resources by fostering competition among providers and improving resource utilization through its decentralized model. This increased accessibility is crucial for making advanced computational capabilities available to a wider audience. Additionally, by leveraging blockchain technology and smart contracts, the GCX ensures secure, transparent, and efficient transactions, which build trust among participants and promote the stability of the marketplace \cite{timofeev2024case, mit2024breakthrough}.

The establishment of the GCX is not just about meeting current computational demands but also about future-proofing the industry. As AI and machine learning technologies continue to evolve, the need for scalable and flexible computational resources will only increase. The GCX is positioned to meet these needs, ensuring that computational power becomes a readily available and manageable commodity, much like electricity or data storage. This will ultimately drive innovation, economic growth, and technological advancement on a global scale \cite{jindal2024compute, dean2020}.

\subsection*{Is Compute the New Oil?}

\subsubsection*{Harnessing Oil Byproducts for the Global Compute Market}

An innovative approach in the compute resource lifecycle is the transformation of oil and its byproducts into “digital oil.” Excess gas from oil extraction sites, often flared due to insufficient transportation infrastructure, can be repurposed to power compute datacenters. By converting this otherwise wasted energy into electricity, these datacenters can generate substantial computational power, contributing significantly to the global compute pool. Additionally, oil itself can directly fuel these sites, providing a primary energy source for computational tasks. This dual utilization not only offers an additional revenue stream for oil companies but also promotes more sustainable use of natural resources. Integrating traditional energy sources with digital infrastructure enhances the efficiency and profitability of both sectors, showcasing a forward-thinking approach to energy and technology.

\subsubsection*{Harnessing Renewable Energy for the Global Compute Market}

Beyond traditional and byproduct energy sources, renewable energy offers a sustainable and efficient means to power the global compute market. Solar, wind, hydroelectric, and geothermal energy can all be harnessed to run compute datacenters, reducing the carbon footprint and promoting environmental sustainability.

Renewable energy provides a reliable and scalable solution to meet the increasing demand for computational power. For instance, solar farms can generate substantial electricity during peak sunlight hours, which can be stored and utilized by compute datacenters. Wind turbines, strategically placed in high-wind areas, can provide consistent energy to power these facilities. Hydroelectric plants, leveraging natural water flow, and geothermal plants, utilizing the Earth’s internal heat, offer continuous energy supplies, ideal for the 24/7 operation of datacenters.

Integrating renewable energy sources into the compute infrastructure reduces dependency on fossil fuels and minimizes greenhouse gas emissions. This not only aligns with global sustainability goals but also enhances the resilience and stability of the compute market by diversifying energy sources. Moreover, renewable energy-powered compute datacenters can attract environmentally conscious businesses and investors, fostering innovation and growth in the green technology sector.

The adoption of renewable energy in the compute market underscores a commitment to creating a more sustainable and eco-friendly digital economy, ensuring that the expansion of computational power does not come at the expense of the environment.

\subsection*{Compute Resource Lifecycle}

The lifecycle of compute resources can be analogized to the "upstream, midstream, and downstream" model used in the oil industry. This analogy helps in understanding the stages from creation to end-use of compute resources.

\subsubsection*{Upstream: Resource Creation and Provisioning}

In the compute resource lifecycle, the upstream stage involves the generation and provision of raw compute resources, akin to oil extraction. Key players in this stage include data centers, cloud service providers, and decentralized compute providers. Data centers are facilities that house the physical servers and infrastructure necessary for generating compute power. Cloud service providers, such as AWS \cite{awsdocs}, Google Cloud \cite{googlecloud}, Azure \cite{azure}, Nexgen Cloud \cite{nexgencloud}, and Oracle \cite{oracle}, offer large-scale compute resources.
Additionally, decentralized compute providers are individuals or smaller entities that contribute spare computing power via decentralized networks.

\subsubsection*{Midstream: Resource Management and Distribution}
The midstream stage involves the aggregation, optimization, and distribution of compute resources, ensuring they are available where and when needed. This stage is analogous to the transportation and storage of oil. Key components include resource scheduling and allocation platforms, task queues and orchestration services, and resource brokerage. Resource scheduling and allocation platforms manage the distribution of compute tasks to available resources, optimizing for efficiency and cost. Task queues and orchestration services handle the queuing and orchestration of compute tasks, ensuring efficient utilization of resources. Resource brokerage involves marketplaces where compute resources are traded, including futures contracts and spot markets.

\subsubsection*{Downstream: Resource Utilization and End-Use}
The downstream stage, similar to refining and retail in the oil industry, involves the use of compute resources for various applications. End-user applications are the final consumers of compute power, such as AI model training, data processing, and running complex simulations. Enterprise solutions are companies that use compute resources for their business operations, including big data analytics, machine learning, and SaaS products. Compute-intensive services rely heavily on compute power and include areas such as video rendering, gaming, and scientific research.

\subsubsection*{Refinery: Compute Optimization and Enhancement}
The refinery stage in the compute resource lifecycle involves optimization services, similar to oil refineries that convert crude oil into usable products. These services enhance raw compute power into more efficient and valuable resources. AI model optimization techniques improve the performance and efficiency of AI models. Performance tuning involves adjusting and optimizing software and hardware settings to maximize compute efficiency. Resource virtualization allows for more efficient use of physical hardware by creating virtual resources.

\section{Learning from Oil to Deliver Compute Resources}

In considering how to effectively deliver compute resources, we draw an analogy to the oil industry, where pipeline hubs or ports serve as key delivery points. Similarly, in the realm of compute resources, establishing an abstraction layer that connects consumers with various providers can act as a delivery hub, simplifying access and distribution, see Fig.~\ref{fig:delivery}.

\begin{figure}[t!]
    \centering
    \includegraphics[width=0.6\linewidth]{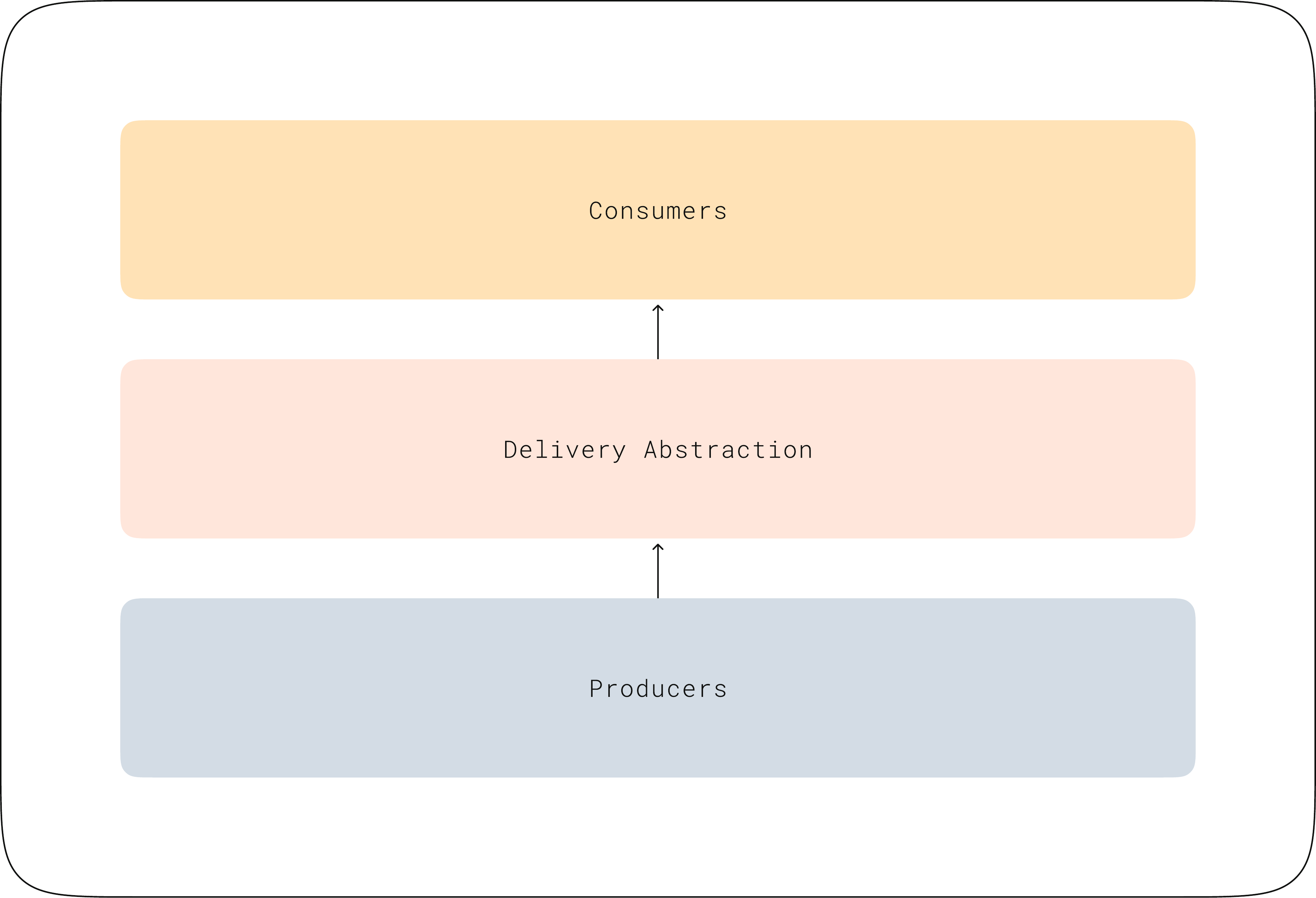}
    \caption[Abstracted compute delivery.]{An abstraction layer facilitates the delivery of compute resources from producers to consumers, streamlining the connection and optimizing the distribution process. This makes the delivery seem like a faucet or a ``delivery API" where consumers, as an oversimplified example, request compute and receive it, making it as easy for a consumer as sending a GET request to a server and a producer to POST \cite{fielding2000architectural}.}
    \label{fig:delivery}
\end{figure}

To create a seamless experience, an abstraction layer, akin to a delivery hub, can facilitate the connection between compute consumers and providers. This layer can be viewed as a ``delivery API" or ``faucet" that manages the flow of compute resources from producers to consumers. This concept is already being explored by companies such as Covalent \cite{covalent2024}, Aethir \cite{aethir2024}, Exabits \cite{exabits2024}, Nosana \cite{nosana2024}, and io.net \cite{ionet2024}, which could be potential partners for future collaboration, or alternatively, we could develop this infrastructure ourselves.

We are working on the assumption that interoperability will be a lot more widespread sooner rather than later. For instance, Microsoft \cite{microsoft2024} is introducing a transformer into their stack that abstracts against CUDA \cite{msmech,azure2023maia,openai2023,nvidia2024}.
This is exemplified by Triton, a project developed by OpenAI \cite{openai}, which aims to provide a more efficient, flexible, and powerful way to implement deep learning models. Triton is designed to achieve high performance without requiring users to write complex CUDA code, thus promoting greater interoperability across various hardware platforms \cite{triton2024}.

\subsection*{Industry Examples}

Several companies exemplify this abstraction layer concept:

\begin{itemize}
    \item Covalent offers a unified API that provides fast and scalable access to historical blockchain data across multiple chains. It simplifies the process of fetching data by maintaining a consistent request and response format, making it easier for developers to build on different blockchains without needing to learn new tools or data structures. This unified approach can be applied to delivering compute resources, ensuring a consistent and efficient user experience.

    \item Aethir specializes in decentralized cloud rendering and compute services, enabling the distribution of high-performance computing power through its decentralized network. This approach reduces reliance on traditional cloud providers and offers an innovative model for delivering compute resources efficiently.

    \item Exabits focuses on providing decentralized cloud services that are robust and scalable. By leveraging a network of distributed nodes, Exabits ensures reliable access to compute power, supporting diverse workloads from AI to large-scale data processing.

    \item Nosana facilitates the distribution of compute resources for CI/CD pipelines \cite{shahin2017continuous} in a decentralized manner. This model helps in optimizing the use of idle compute power, making it accessible to developers globally and enhancing the efficiency of continuous integration and deployment processes.

    \item io.net provides a decentralized computing network that allows machine learning engineers to access scalable distributed clusters at a fraction of the cost of traditional centralized services. By offering instant and permissionless access to global GPU resources, io.net ensures flexibility, speed, and affordability. This network supports a wide range of GPUs and CPUs \cite{stallings2003computer}, including advanced Apple silicon chips \cite{apple2020silicon}, making it a versatile solution for diverse computing needs.
\end{itemize}

Implementing an abstraction layer for delivering compute resources offers several notable advantages, each contributing to the overall effectiveness and efficiency of the system. Flexibility is one of the primary benefits of an abstraction layer. Users can access compute resources from various providers without being locked into a specific vendor. This flexibility allows organizations to choose the most suitable resources for their specific needs, whether it's due to cost considerations, performance requirements, or specific technical features offered by different providers. By decoupling the compute resource from the provider, the system can seamlessly integrate with new technologies and vendors as they become available, ensuring that users always have access to the best possible resources.

Scalability is another significant advantage. An abstraction layer enables the system to dynamically scale based on demand. This means that as the need for compute resources increases or decreases, the system can adjust accordingly, provisioning additional resources during peak times and deallocating them when they are no longer needed. This dynamic scaling ensures that compute resources are always available when required, preventing bottlenecks and ensuring smooth operation even during high-demand periods. Moreover, this scalability can lead to cost savings, as resources are only used when necessary, avoiding the expense of maintaining idle capacity.

Efficiency in the allocation and utilization of compute power is also greatly enhanced by an abstraction layer. By optimizing how resources are allocated, the system can improve overall efficiency, reducing both costs and environmental impact. Efficient resource allocation means that fewer resources are wasted, and the compute power is used more effectively, leading to better performance and lower operational costs. Additionally, this optimization can help in reducing the carbon footprint of computing operations, as more efficient use of resources typically translates to lower energy consumption.

Incorporating an abstraction layer for delivering compute resources aligns with the principles of decentralization and efficiency. By drawing inspiration from existing models in the blockchain and decentralized compute industries, we can create a robust and scalable infrastructure that meets the growing demands for high-performance computing. This approach not only addresses current challenges but also positions us to leverage future advancements in decentralized technologies, ensuring that our solution remains relevant and competitive.
\section{Learning from Power to Enhance Compute Resource Trading}

This section expands on the comparative analysis of the power and computational resource markets by specifically focusing on the metrics used in each market. This detailed comparison aims to highlight how these metrics influence market operations and the potential for adopting analogous measures across markets to improve performance and stability.

We note also that, in essence, the drive towards the commodification of computing resources is primarily motivated by the need to enhance insights into future cash flows and mitigate the risks associated with uncertain future utilization levels. It's less about the actual computing power and more about providing financial predictability and stability. Data centers are becoming the new power plants in this evolving landscape \cite{cohen2013compute}.

Both the compute resource market and the power market share several similarities, such as the need for real-time balancing, price volatility, and the importance of reliable delivery. However, they also have key differences. Unlike power, compute resources can be queued and scheduled, offering opportunities for innovative efficiency optimizations not possible in the power market \cite{balatsky2015, baliga2010energy}. By adopting and adapting methodologies from the power market, we can enhance the operational efficiency, reliability, and sustainability of the compute resource market \cite{hinton2023future}. In turn, this supports better clarity for power infrastructure builders and their planning, as the optimized scheduling and queuing of compute resources can lead to more predictable and manageable power demands.

\subsection*{Power Market Metrics}

In the power market, the following metrics play a key role in maintaining efficiency, reliability, and sustainability:

\begin{itemize}
    \item \textbf{Watt-hours (Wh)}: Measures the quantity of power used, serving as the fundamental unit for billing and trading.
    \item \textbf{Frequency and Voltage}: Critical for ensuring the stability and quality of the electrical supply.
    \item \textbf{System Average Interruption Duration Index (SAIDI) and System Average Interruption Frequency Index (SAIFI)}: Assess the reliability and availability of the power supply.
    \item \textbf{Power Factor}: Indicates the efficiency of power use, affecting how much charge users incur.
    \item \textbf{Emissions Intensity}: Measures the environmental impact per unit of power generated, increasingly relevant in carbon trading schemes.
\end{itemize}

\begin{figure}[t!]
\captionsetup{name=Table}
\setcounter{figure}{1}
\captionsetup{list=off} 
\centering
\includegraphics[width=\linewidth]{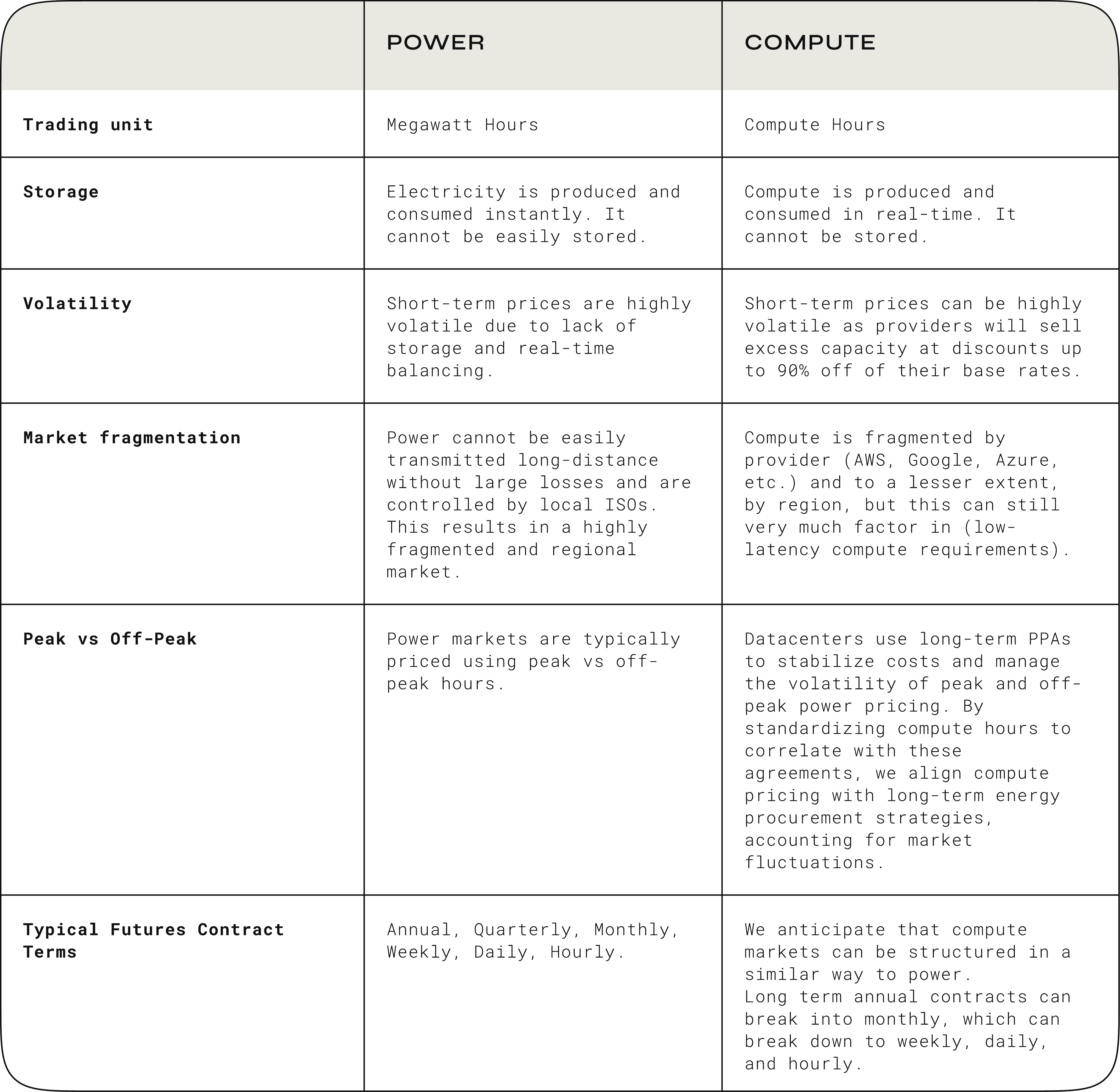}
\caption[Comparison of Power and Compute Markets.]{Comparison of Power and Compute Markets: This table outlines the key differences between power and compute markets, including trading units, storage, volatility (see \cite{AWS_Spot,Google_Preemptible,Azure_Spot} for the reference to 90\% spot discount), market fragmentation, peak vs. off-peak pricing (PPA is a Power Purchase Agreement \cite{mendicino2019corporate}), and typical futures contract terms.}
\label{tab:power_compute_comparison}
\end{figure}

\subsection*{Compute Resource Market Metrics}
Similarly, we propose the following metrics for the compute market:
\begin{itemize}
\item \textbf{Compute Hours}: Analogous to Wh, measuring the quantity of computational work done.
\item \textbf{Latency and Throughput}: Key performance metrics that determine the efficiency and speed of data processing.
\item \textbf{Mean Time Between Failures (MTBF) and Mean Time to Repair (MTTR)}: Indicators of system reliability and service continuity, similar to SAIDI/SAIFI \cite{cepin2011reliability}.
\item \textbf{Utilization Efficiency}: Measures how effectively compute resources are being used, akin to the power factor in electricity.
\item \textbf{Energy Efficiency}: Reflects the power consumption per unit of compute work, parallel to emissions intensity in highlighting sustainability.
\end{itemize}

Tab.~(\ref{tab:power_compute_comparison}) provides a side-by-side comparison of key characteristics between power and compute markets.

Both markets utilize metrics that measure quantity (Wh and Compute Hours), efficiency (Power Factor and Utilization Efficiency), reliability (SAIDI/SAIFI and MTBF/MTTR), and environmental impact (Emissions Intensity and Energy Efficiency).
These parallels suggest that methodologies for optimizing these metrics in one market could be adapted for the other.

While both markets share similarities in metric goals, the physical nature of power demands real-time balance and immediate consumption, posing unique challenges not faced in the compute market.
Conversely, the compute market’s flexibility in queuing and scheduling tasks offers opportunities for innovative efficiency optimizations not possible in power markets. This flexibility not only enhances compute market operations but also supports power infrastructure planning by providing clearer and more predictable power requirements.

The detailed examination of market metrics reveals significant opportunities for mutual learning between the power and compute resource markets.
By adopting and adapting methodologies and metrics from the power market, the compute resource market can enhance its operational efficiency, reliability, and environmental sustainability, fostering a more robust and equitable marketplace \cite{balatsky2015, baliga2010energy, hinton2023future}.
\section{Standardizing Compute Units}

To standardize compute units, we propose the Compute Hour (CH), which abstracts the complexities of different hardware architectures and provides a consistent measure of computational work; this defines the compute commodity.
Towards this, defining a reference system is a foundational step in this process. The reference system should have a balanced performance profile, encompassing CPU, GPU, memory, and storage capabilities. Benchmarking this system involves running a suite of standardized tests to measure performance across various dimensions, ensuring fair and accurate trading of compute resources \cite{ren2009zero, xu2017efficient}.

Also, it is difficult to use a compute cycle itself as a fundamental unit. A compute cycle represents a single instruction period of a computer's CPU or GPU. One cycle is the time it takes for a CPU or GPU to fetch an instruction, decode it, execute it, and store the result. On the surface, measuring compute resources by counting the number of compute cycles seems straightforward—similar to counting revolutions on a power meter for kWh. This approach suggests a direct measurement of ``work done" by the computing systems.

However, different CPUs and GPUs have different architectures (e.g., x86 vs. ARM for CPUs, and NVIDIA vs. AMD for GPUs), and their cycles cannot be directly compared. What one CPU or GPU can achieve in a single cycle might take another multiple cycles, or vice versa. Different CPUs and GPUs operate at different clock speeds, meaning the number of cycles per second can vary widely. Thus, compute cycles alone do not provide a uniform standard of measurement across different hardware.

Modern CPUs and GPUs can execute multiple instructions per cycle through techniques like pipelining and parallel execution. Therefore, the efficiency of cycle usage can vary, complicating a direct comparison based solely on cycle count. Different CPUs and GPUs support different sets of instructions (Instruction Set Architecture), and some can achieve more per cycle than others depending on the instruction mix. Operating systems and software optimizations play a significant role in how efficiently these cycles are used.

The CH abstracts away these complexities and provides a consistent measure of computational work. This ensures fair and accurate trading of compute resources in the marketplace, regardless of the underlying hardware differences:

\begin{eqnarray}
\text{Compute Hours (CH)} = \frac{\text{Performance (FLOPS)}}{\text{Reference Performance (FLOPS)}} \times \text{Operational Hours}  \nonumber \\ \label{eqn:ch}
\end{eqnarray}

Where Performance (FLOPS) is the actual performance output of the compute system measured in FLOPS.
Reference Performance (FLOPS) is the benchmark performance of the reference system, also measured in FLOPS, and Operational Hours, or physical/wall-clock hours, is the duration in hours for which the system is utilized for computing tasks.
FLOPS is floating-point operations per second, a measure of computational performance, especially in tasks involving real-number calculations common in AI and machine learning.

This formula allows for the normalization of computing performance across different systems and configurations. By using a reference system as a baseline, the Compute Hours for any system can be calculated by scaling its performance relative to this standard benchmark. This scaling ensures that a Compute Hour represents the same amount of computational work, irrespective of the underlying hardware differences.

Our goal is to commodify compute resources by creating a universal metric that can be applied across a wide range of computational tasks. This doesn't mean that one reference system applies to all tasks uniformly. Instead, it means that we have a standard approach to measure and compare computational work by normalizing against task-specific reference systems.
In essence, the Compute Hours metric is intended to provide a standardized measure of computational work that can be applied broadly, while still respecting the unique characteristics of different types of tasks. 

\subsection*{Estimating Compute Hours for AI Models}

Mapping the size of an AI training model to the amount of compute time needed can be approached by understanding the relationship between the model’s architecture, the dataset, and the computational resources used. To learn more about AI models and their architecture, please consult, for example Refs.~\cite{paperspace2020,brownlee2018,deepai2021,pytorch2021}.
What follows is a comprehensive way to estimate the compute time based on model size. We assume the model architecture is known in detail, but this does not affect the generality of the result.

Consider a Convolutional Neural Network (CNN). The CNN processes an input image through several layers, each performing specific operations. Fig.~(\ref{fig:CNN_architecture}) illustrates a typical CNN architecture used for image classification.

\begin{figure}[t!]
    \setcounter{figure}{4}
    \centering
    \includegraphics[width=\linewidth]{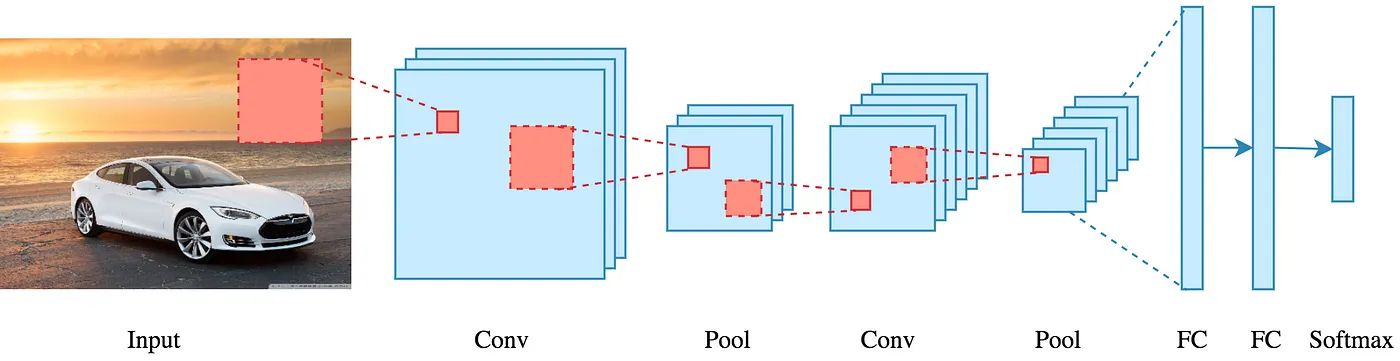}
    \caption[Diagram of a Convolutional Neural Network (CNN) architecture.]{Diagram of a Convolutional Neural Network (CNN) architecture. The image input is processed through multiple layers including convolutional (Conv) and pooling (Pool) layers. The final layers consist of fully connected (FC) layers and a softmax function for classification. This architecture illustrates the components used to calculate the Floating Point Operations (FLOPs) for each layer, necessary for estimating compute hours based on model architecture \cite{dertad2017}.}
    \label{fig:CNN_architecture}
\end{figure}

\begin{itemize}
    \item \textbf{Input}: The raw input image fed into the network.
    \item \textbf{Conv}: Convolutional layers that apply filters to extract features from the input image.
    \item \textbf{Pool}: Pooling layers that reduce the spatial dimensions of the feature maps, retaining important information while reducing computational load.
    \item \textbf{FC}: Fully connected layers that combine the features learned by convolutional and pooling layers to make final predictions.
    \item \textbf{Softmax}: The output layer that provides the probability distribution over classes.
\end{itemize}

In what follows, we will consider a general use-case, but the idea is that this can be applied to specific AI models like the one in Fig.~(\ref{fig:CNN_architecture}).

\subsubsection*{Understand Model Architecture}
   \begin{itemize}
      \item \textbf{Layers and Parameters}: The number of layers, types of layers (e.g., convolutional, fully connected), and the number of parameters in each layer.
      \item \textbf{Operations}: Types of operations involved, such as matrix multiplications, convolutions, and activations.
   \end{itemize}

\subsubsection*{Calculate Floating Point Operations (FLOPs\footnote{FLOPS stands for floating-point operations per second, indicating a rate of computation. In contrast, FLOPs is the plural form of floating-point operation, referring to the number of operations. 1 TFLOP is a single teraFLOP which is $10^{12}$ FLOPs.})}

   \begin{itemize}
      \item \textbf{Per Layer FLOPs}: Calculate the number of FLOPs for each layer. For example, a convolutional layer’s FLOPs can be estimated using:

      \begin{eqnarray}
      \text{FLOPs per layer} &= 2 \times \text{input height} \times \text{input width} \times \text{input channels} \times \text{output channels} \nonumber\\ &\times \text{kernel height} \times \text{kernel width} \times \text{output height} \times \text{output width} \nonumber
      \end{eqnarray}
      \item \textbf{Total FLOPs}: Sum the FLOPs for all layers to get the total FLOPs required for one forward pass.
   \end{itemize}
   
\subsubsection*{Training Iterations}
   \begin{itemize}
      \item \textbf{Epochs}: Determine the number of epochs (complete passes through the training dataset).
      \item \textbf{Batch Size}: Number of samples processed at once.
      \item \textbf{Iterations}: Calculate the total number of iterations as \(\text{total samples} / \text{batch size}\).
   \end{itemize}

\subsubsection*{Example Calculation}

For a given model, suppose:
\begin{itemize}
    \item \textbf{Model FLOP}: 1 trillion FLOPs (1 TFLOP) per forward pass
    \item \textbf{Dataset Size}: 1 million samples
    \item \textbf{Batch Size}: 256
    \item \textbf{Epochs}: 10
\end{itemize}

\begin{eqnarray}
\text{Total FLOPs Needed} &=& 10^{12} \, \text{Model FLOPs} \times \left(\frac{10^6 \, \text{samples}}{256 \, \text{batch size}}\right) \times 10 \, \text{epochs} \nonumber \\
&=& 3.9 \times 10^{16} \, \text{FLOPs} = 39 \, \text{PFLOPs}
\end{eqnarray}

By specifying further over which duration this needs to be completed we can obtain a rate: This is ``Performance (FLOPS)" in Eq.~(\ref{eqn:ch}); the amount of CH follows from that.
The user can now buy this from the GCX when also specifying when the computation needs to take place.
To be sure, the concept of Compute Hour can be generalized beyond AI.

Finally, let us say the AI model details are not known. We can still estimate the total number of FLOPs by using a reference GPU. This is something that can be provided on the GCX as well.
This method allows us to guide users to the appropriate market for trading and provides valuable recommendations based on the computational requirements.

\subsection{Benchmarking and Normalization}

As part of defining the Compute Hours we introduced a reference system. This is a foundational step in standardizing compute units and facilitating the commodification of compute resources. By leveraging established benchmarking practices from the supercomputing industry \cite{marathe2013comparative,inadomi2015analyzing}, we can create a robust and credible framework for measuring and comparing computational power. This framework ensures that all compute resources are evaluated on a consistent and fair basis, promoting transparency and efficiency in the compute marketplace.

Benchmarking involves running a suite of standardized tests to measure a system's performance across various dimensions, such as processing speed, memory bandwidth, and input/output operations. The results of these tests are then compared to the reference configuration to determine the equivalent number of Compute Hours.

For example, if a high-performance server is twice as fast as the reference configuration in executing the benchmark suite, it would be equivalent to two Compute Hours per physical hour of operation. Conversely, a less powerful system that achieves half the benchmark performance would be rated at 0.5 Compute Hours per physical hour.

\subsubsection*{Key Areas for Universal Benchmark Consensus}

\subsubsection*{Performance Metrics}
To standardize computational performance, we propose using established benchmarks such as HPL (High Performance Linpack), SPEC CPU, and real-world application benchmarks. For instance, implementing the HPL benchmark can effectively measure the floating-point performance of a supercomputer cluster. Key contributions in this area come from several researchers. The authors in \cite{dongarra2003linpack,hoefler2009message,shalf2010exascale,bader2009stinger} have significantly contributed through the TOP500 and HPL benchmarks, been influential in establishing MPI \cite{gropp1999using} standards and benchmarks, worked on performance analysis in exascale computing providing valuable insights, and advancing the field on multithreaded algorithm performance.

\subsubsection*{Energy Efficiency}
Energy efficiency is critical for optimizing computational resource utilization. Measuring power consumption under various loads using tools like PowerAPI \cite{garnier2013powerapi} and integrating energy-saving techniques such as dynamic voltage and frequency scaling (DVFS) are essential steps. For example, DVFS can optimize energy consumption during periods of low usage. Here, energy-efficient algorithms exist \cite{dongarra2003linpack}, while the work of \cite{matsuoka2007tsubame} in energy-efficient supercomputing is well-regarded. Research on energy efficiency with DVS also provides significant insights \cite{rountree2012energy,simunic2001dynamic}.

\subsubsection*{Scalability}
Scalability is crucial for accommodating the growing demand for computational power across heterogeneous systems. Developing benchmarks that test scalability by combining CPUs, GPUs, and accelerators is essential. A mix of LINPACK and custom benchmarks can evaluate performance across different architectures and configurations. Work on scalable supercomputing architectures from \cite{matsuoka2007tsubame}, contributions to scalability in exascale architectures from \cite{shalf2010exascale}, and research on scalable communication patterns from \cite{hoefler2009message} are pivotal in this domain.

\subsubsection*{Portability}
Ensuring that benchmarks can run on various platforms (e.g., x86 \cite{shvets2010x86}, ARM \cite{furber2000arm}, GPUs \cite{owens2008gpu}) without modification is fundamental for a universal compute benchmark. Testing benchmarks on different cloud providers and on-premise hardware ensures consistent performance metrics. Here, the work of \cite{dongarra2003linpack} on the portability of LAPACK and BLAS libraries, contributions to portability in TSUBAME \cite{matsuoka2008tsubame} systems, and research on portability in MPI applications \cite{hoefler2009message,gropp1999using} are key references in this area.

\subsubsection*{Fault Tolerance}
Integrating fault injection and recovery tests into the benchmarking suite is essential to measure system resilience. Tools like Chaos Monkey \cite{chaosmonkey} can simulate failures and measure a system's ability to recover and maintain performance.

\subsubsection*{Cost-effectiveness}
Developing economic models that compare performance-per-dollar across different configurations is vital for identifying cost-effective solutions. Using real-world cost data from cloud providers and hardware vendors, we can calculate the total cost of ownership and return on investment for various compute setups.
Here, \cite{bader2009stinger} has research on cost models for graph computing and \cite{wahib2013high} has done work on cost-effective HPC/AI solutions which are instrumental in developing these economic models \cite{li2010hybrid}.

\subsection{Defining the Reference System}

The establishment of a reference system is pivotal in standardizing compute units and ensuring fair benchmarking across different compute resources. A reference system serves as a benchmark against which the performance of other systems is measured. This process is akin to how supercomputers are benchmarked to assess their performance.

\subsubsection*{Criteria for the Reference System}

The reference system must be a well-documented and widely recognized configuration to ensure credibility and consistency. The following criteria are considered when defining the reference system:

\begin{itemize}
    \item \textbf{Performance:} The reference system should have a balanced and representative performance profile, encompassing CPU, GPU, memory, and storage capabilities.
    \item \textbf{Availability:} It should be based on commercially available hardware to ensure that the benchmarks are reproducible.
    \item \textbf{Documentation:} Comprehensive documentation of the hardware and software configurations, including operating systems and drivers, must be available.
    \item \textbf{Scalability:} The system should be scalable to allow for adjustments in benchmarking as technology advances.
\end{itemize}

In the supercomputing world, benchmarks like the HPL are used to evaluate and rank supercomputers \cite{dongarra2003high}. Additional benchmarks like HPCG (High Performance Conjugate Gradients) \cite{dongarra2003high}, STREAM (memory bandwidth) \cite{mccalpin1995stream}, and SPEC CPU (single-thread performance) \cite{spec2006cpu} can also be used for specific aspects of performance.
The HPL benchmark measures a system's floating-point computing power and is the basis for the TOP500 list of the world's most powerful supercomputers \cite{top500}. Similarly, we propose using a standardized benchmark suite to evaluate and define the reference system.

\subsubsection*{Benchmarking Process}

Benchmarking GPUs to define a compute hour involves measuring the performance of the GPU in terms of computational tasks over a specific period. The first step is to define the benchmarking metrics. These include FLOPS (Floating Point Operations Per Second), which measures the raw computational power; throughput, which measures the number of tasks or operations the GPU can handle per unit of time; latency, which measures the time taken to complete a single operation or task; power consumption, which measures the energy efficiency of the GPU; and memory bandwidth, which measures the speed at which data can be read from or written to the GPU’s memory.

Next, select benchmarking tools and software. Established tools such as SPECviewperf \cite{specviewperf}, Geekbench \cite{geekbench}, 3DMark \cite{3dmark}, GPGPU \cite{owens2008gpu} Benchmarks like Rodinia \cite{rodinia} or Parboil \cite{parboil}, and Deep Learning \cite{lecun2015deep} Benchmarks such as MLPerf \cite{mlperf} are commonly used. Set up the testing environment, ensuring the GPU is installed in a standard configuration that matches common usage scenarios and that the necessary drivers, benchmarking tools, and specific software libraries are installed.

Run synthetic benchmarks to simulate various computational tasks, which could include matrix multiplications, deep learning model training, rendering tasks, or other GPU-intensive operations. Tools like CUDA-Z \cite{cudaz}, OpenCL Benchmark \cite{opencl_benchmark}, Blender \cite{blender} for rendering tests, and TensorFlow/PyTorch \cite{tensorflow, pytorch2021} for deep learning tasks are useful for this purpose. In addition to synthetic benchmarks, run real-world benchmarks using applications relevant to the target use case, such as Blender for rendering, Adobe Premiere Pro \cite{helpx2023premiere} for video editing, and TensorFlow/PyTorch for machine learning.

Collect and analyze the performance data, measuring metrics such as FLOPS, task completion time, throughput, and power consumption during the benchmarks. Monitoring tools like NVIDIA-SMI \cite{nvidiasmi} can log performance data. Normalize the results to standardize the compute hour definition.
For example, if a GPU completes a certain benchmark task in 30 minutes, it would have utilized 0.5 compute hours if the task is defined as a 1 compute hour benchmark. Aggregate the performance data to define a compute hour for the GPU, based on the average performance across different benchmarks or a weighted score that emphasizes specific tasks.

Examples of existing processes for benchmarking include MLPerf, which provides a suite of benchmarks for evaluating machine learning performance, including training and inference benchmarks for various neural network models. SPECviewperf benchmarks graphics performance by running a series of tests using real-world applications, measuring the ability of GPUs to handle 3D graphics workloads. Geekbench \cite{geekbench} offers a cross-platform benchmark that measures CPU and GPU performance, including tests for various compute tasks.
The software 3DMark \cite{3dmark} is widely used for benchmarking gaming performance, running a series of graphics and computational tests to evaluate the capabilities of GPUs in gaming scenarios.

\begin{itemize}
\item \textbf{Setup}: Install an NVIDIA RTX 3080 \cite{nvidia_rtx_3080} in a test machine with the latest drivers. Set up TensorFlow with CUDA and cuDNN \cite{chetlur2014cudnn}.
\item \textbf{Synthetic Benchmark}: Run CUDA-Z to measure FLOPS. Execute TensorFlow benchmarks with a ResNet-50 model \cite{Koonce2021} to measure training time and throughput.
\item \textbf{Real-World Benchmark}: Use Blender to render a high-resolution 3D scene. Run inference benchmarks with TensorFlow using a pre-trained BERT model \cite{devlin2019bert}.
\item \textbf{Data Collection}: Record the time taken for each task. Measure power consumption using NVIDIA-SMI. Log memory bandwidth utilization during the tasks.
\item \textbf{Analysis}: Calculate the average performance metrics. Normalize the results to define the compute hour based on task completion times and energy efficiency.
\item \textbf{Report}: Publish the normalized results, defining the compute hour for the RTX 3080 based on the aggregated performance data.
\end{itemize}

By following these steps and leveraging existing benchmarking tools and methods, the compute hour for a GPU can be accurately defined and standardized, allowing for fair and transparent trading in platforms like the GCX.

\subsubsection*{Normalization and Conversion}

Once the reference system's performance metrics are established, other compute resources can be normalized to this standard. The normalization process involves comparing a system's benchmark results to those of the reference system. The formula for calculating Compute Hours, as previously discussed, is applied to perform this conversion.

For example, if the reference system achieves 1 TFLOP in the benchmark test, and another system achieves 0.5 TFLOPs, the latter would be assigned 0.5 Compute Hours for every physical hour of operation. Conversely, a system achieving 2 TFLOPs would be assigned 2 Compute Hours per physical hour.

\subsubsection*{Updating the Reference System}

As technology evolves, it is crucial to periodically update the reference system to reflect advancements in computational hardware and software. Regular updates ensure that the benchmarking process remains relevant and accurate. The criteria for updates include:

\begin{itemize}
    \item \textbf{Technological Advancements:} Incorporating new hardware and software innovations.
    \item \textbf{Market Trends:} Reflecting shifts in the computational resource market.
    \item \textbf{Feedback from Stakeholders:} Considering input from compute providers and users.
\end{itemize}

By maintaining an up-to-date reference system, we ensure that the commodification of compute resources remains fair, transparent, and aligned with current technological standards.

\subsection{Grading Compute Resources and ESG}

Compute resources can vary widely in terms of performance, reliability, and suitability for different tasks.
To facilitate transparent and efficient trading, we will grade compute resources based on their benchmark performance and additional criteria such as uptime, reliability, and energy efficiency.

Performance grades categorize compute resources into tiers based on their benchmark results. For instance:
\begin{itemize}
    \item \textbf{Grade A:} High-performance systems with benchmark results significantly above the reference configuration.
    \item \textbf{Grade B:} Systems with performance moderately above the reference configuration.
    \item \textbf{Grade C:} Systems with performance close to the reference configuration.
    \item \textbf{Grade D:} Systems with performance below the reference configuration.
\end{itemize}

Reliability is crucial for many computational tasks. Compute resources will be graded on their historical uptime and failure rates. For instance:
\begin{itemize}
    \item \textbf{Grade 1:} Systems with uptime exceeding 99.9\%.
    \item \textbf{Grade 2:} Systems with uptime between 99.0\% and 99.9\%.
    \item \textbf{Grade 3:} Systems with uptime between 95.0\% and 99.0\%.
    \item \textbf{Grade 4:} Systems with uptime below 95.0\%.
\end{itemize}

Energy efficiency is increasingly important in modern computing \cite{baliga2010}. Compute resources will be graded based on their energy consumption relative to their performance. For instance:

\begin{itemize}
    \item \textbf{Grade X:} Highly energy-efficient systems with low power consumption per Compute Hour.
    \item \textbf{Grade Y:} Moderately energy-efficient systems.
    \item \textbf{Grade Z:} Systems with higher energy consumption.
\end{itemize}

Implementing a grading system for compute resources based on performance, reliability, and energy efficiency can significantly promote the use of ESG (Environmental, Social, and Governance)-friendly sources of compute. By providing clear and transparent benchmarks, users can make informed decisions about the sustainability and efficiency of their computational resources. For instance, high-performance grades (e.g., Grade A) combined with high energy efficiency grades (e.g., Grade X) will enable users to select top-tier compute resources that not only meet their performance needs but also align with their environmental sustainability goals.
The price point difference will encourages data centers and compute providers to adopt greener technologies and practices to achieve higher grades, thereby reducing the carbon footprint of computational activities. Ultimately, this fosters a market where environmental responsibility and efficiency are rewarded, driving innovation and sustainability in the computing industry.
\section{Blockchain-based Clearing and Guarantee Pools in the GCX}

At the heart of the GCX lies the innovative concept of clearing and guarantee pools, where participants and guarantors play a crucial role in ensuring the integrity and efficiency of the compute market. Critical to a well-functioning commodities futures market are the following elements:

\begin{enumerate}
    \item \textbf{Convergence of Futures Price to Spot Price}: At expiration, the futures price must converge to the actual spot price of the underlying commodity. This is typically accomplished by ensuring that holders of long positions in futures contracts can take delivery of the underlying asset, while short position holders are obligated to deliver the asset as per the contract terms.
    
    \item \textbf{Adaptive Risk/Margin System}: A well-defined risk/margin system that can adapt to changing market conditions is essential. Mechanisms must be in place to ensure that sufficient margin is held by market participants.
    
    \item \textbf{Margin Maintenance and Liquidation Mechanisms}: Ensuring that market participants maintain sufficient margin and providing mechanisms to liquidate positions that fail to meet margin requirements are crucial for market stability.
    
    \item \textbf{Backstop Liquidity Pool}: A liquidity pool that can cover losses in case liquidated accounts do not have sufficient funds to meet their obligations is crucial. This pool acts as a safety net to ensure market stability.
\end{enumerate}

In traditional financial markets, points 1 and 2 have long been established by major exchanges. However, the industry is transitioning from older portfolio risk models (e.g., CME SPAN) to more modern Value at Risk (VaR) methods. For the GCX, we have the opportunity to build a margin system based on the latest advances in risk management techniques, unencumbered by legacy systems.

Regarding points 3 and 4, these functions were traditionally provided by clearing members at major exchanges. Non-clearing market participants must have their trades guaranteed by a clearing member, who is ultimately responsible for ensuring the financial obligations of its customers. Clearing members must also keep customer funds segregated from their own.

In the context of the GCX, Guarantee Pools can assume the role of traditional clearing firms by ensuring adequate margin is kept by their customers, that such margin is segregated from the pool’s funds, and that liquidations can be enforced. Blockchain technology can facilitate this process.

Additionally, an Insurance Fund can serve as a final backstop against any financial losses resulting from a pool being unable to meet its obligations. All pools would be required to stake tokens to the Insurance Fund in proportion to their overall risk. The exchange itself (GCX) would also stake tokens into the Insurance Fund to align all parties’ interests in maintaining proper risk controls. Furthermore, outside participants can contribute tokens to the Insurance Fund in return for yield, enhancing the pool’s robustness and providing additional incentives for risk management.

\subsection{Layered Architecture of the GCX}

The GCX is structured into several distinct layers, each with specific roles and responsibilities. This multi-layered architecture ensures a robust and efficient operation of the compute market, leveraging blockchain technology for enhanced transparency and security. The layers and their interactions are illustrated in Fig.~(\ref{fig:gcx_diagram}).

\subsubsection{Market Layer}

The Market Layer comprises the participants who interact with the GCX to trade compute resources. These participants include:

\begin{itemize}
    \item \textbf{Hedgers}: Entities seeking to hedge against price volatility in compute resources.
    \item \textbf{Traders}: Participants who engage in buying and selling compute resources to profit from price movements.
    \item \textbf{Short Sellers}: Entities that sell compute resources they do not currently own, betting on future price declines.
    \item \textbf{Market Makers}: Entities that provide liquidity to the market by continuously quoting buy and sell prices. These market makers, leveraging their high trading volumes, may opt to purchase GCX tokens to benefit from reduced transaction fees.
\end{itemize}

\subsubsection{App Layer}

The App Layer consists of applications developed by various companies, including the GCX itself. These apps provide user interfaces and frontends for interacting with the GCX platform. Key components include:

\begin{itemize}
    \item \textbf{GCX's App}: The primary application developed by the GCX for trading compute resources.
    \item \textbf{Company 2's App}: An example of a third-party application integrated with the GCX.
    \item \textbf{Company 3's App}: Another example of a third-party application integrated with the GCX.
    \item \textbf{User Interface \& Frontend}: Interfaces that allow users to interact with the apps and manage their trading activities.
    \item \textbf{API \& SDK (Application Programming Interface and Software Development Kit)}: A set of protocols, tools, and libraries provided by GCX that allows developers to create and integrate applications seamlessly with the GCX platform. This enables faster development and integration of third-party apps.
\end{itemize}

\subsubsection{Clearing Layer}

The Clearing Layer facilitates smooth and secure trade settlements. It includes guarantors who provide trade guarantees and manage the delivery of compute resources. Guarantors also serve as gatekeepers to the GCX, ensuring market participants maintain sufficient collateral to cover their risk both pre- and post-trade.

To address potential concerns regarding the requirement for compute providers to post collateral, it is important to highlight the significant benefits of this approach. Requiring collateral helps mitigate counterparty risk, ensuring that market participants fulfill their obligations, thereby enhancing market stability and integrity. This requirement increases buyer confidence, fostering trust and encouraging market participation. While posting collateral introduces an upfront cost for compute providers, it standardizes risk management practices, leading to more efficient operations. Additionally, collateral enables larger and more complex transactions, promoting market growth and higher profitability for compute providers. Though the collateral requirement is a downside, it enables the market to function effectively, resulting in a net positive for participants. Therefore, the collateral requirement strikes a balance between ensuring market integrity and providing a secure platform for compute providers, ultimately leading to long-term benefits that outweigh the initial costs.
Key components include:

\begin{itemize}
    \item \textbf{Guarantor 1, 2, 3}: Entities that guarantee the fulfillment of trades. Guarantors may be subject to losses of their own collateral if a customer's collateral is insufficient to cover losses in case of default. This incentivizes Guarantors to maintain proper risk controls over their customers.
    \item \textbf{Proof of Compute/Capacity (e.g., via DePIN + Truebit)}: Mechanisms to verify the availability and delivery of compute resources.
    \item \textbf{Delivery of Compute}: The process of delivering compute resources as per the contract terms.
\end{itemize}

\subsubsection{Risk Management Layer}

The Risk Management Layer monitors and manages the risks associated with trading on the GCX. It includes:

\begin{itemize}
    \item \textbf{APIs \& SDKs}: Tools for developers to integrate their app into the GCX and to integrate risk management features into their applications as well.
    \item \textbf{Risk Engine: Monitoring and Analytics}: Systems that analyze and monitor risk factors to ensure market stability. The risk engine also scores the participants and ensures the entire health of the platform.
\end{itemize}

\subsubsection{Exchange Layer (Offchain)}

The Exchange Layer manages the off-chain operations of the GCX, including matching buy and sell orders and executing trades. It is also responsible for determining which products and derivatives are listed and establishing risk calculation methods.
Key components include:

\begin{itemize}
    \item \textbf{Global Compute Exchange (GCX)}: The core exchange platform for trading compute resources.
    \item \textbf{Marketplaces (Spot, Derivatives: Futures, Options)}: Various marketplaces within the GCX for trading different types of compute contracts.
\end{itemize}

\subsubsection{Blockchain Layer (Onchain)}

The Blockchain Layer manages the onchain operations, leveraging blockchain technology for transparency and security. It includes:

\begin{itemize}
    \item \textbf{Smart Contract Ecosystem}: A system of smart contracts that govern the operations of the GCX.
    \item \textbf{Token Interface (GCX Token)}: The primary currency/utility token used within the GCX ecosystem.
    \item \textbf{Token Collateral}: Tokens used as collateral for trades, including collateral tokens (USDC, USDT) and proof of compute tokens.
    \item \textbf{Customer Collateral (Segregated)}: Collateral provided by customers, kept separate from the pool’s funds.
    \item \textbf{Guarantor Pool Collateral, GCX tokens, Proof of Compute}: Collateral provided by guarantors to back their guarantees.
    \item \textbf{Insurance Pool}: A pool of funds to cover losses in case of defaults.
    \item \textbf{Staking and Slashing}: Mechanisms to ensure compliance and penalize non-performance.
    \item \textbf{Yield Aggregation/Distribution}: Systems to distribute earnings to participants.
    \item \textbf{Liquidation Engine}: Systems to liquidate positions that fail to meet margin requirements.
    \item \textbf{Connectors to broader DeFi Ecosystem}: Integrations with the broader decentralized finance ecosystem for additional liquidity and services.
\end{itemize}

We discuss the utility token in more detail in Sec.~(\ref{sec:token}).

\subsection{Process Flow: Clearing and Guarantee Pools}

The process flow for clearing and guarantee pools within the GCX involves multiple steps with the goal to ensure that the platform operates smoothly and securely.

\begin{figure}[t!]
    \centering
    \includegraphics[width=0.64\linewidth]{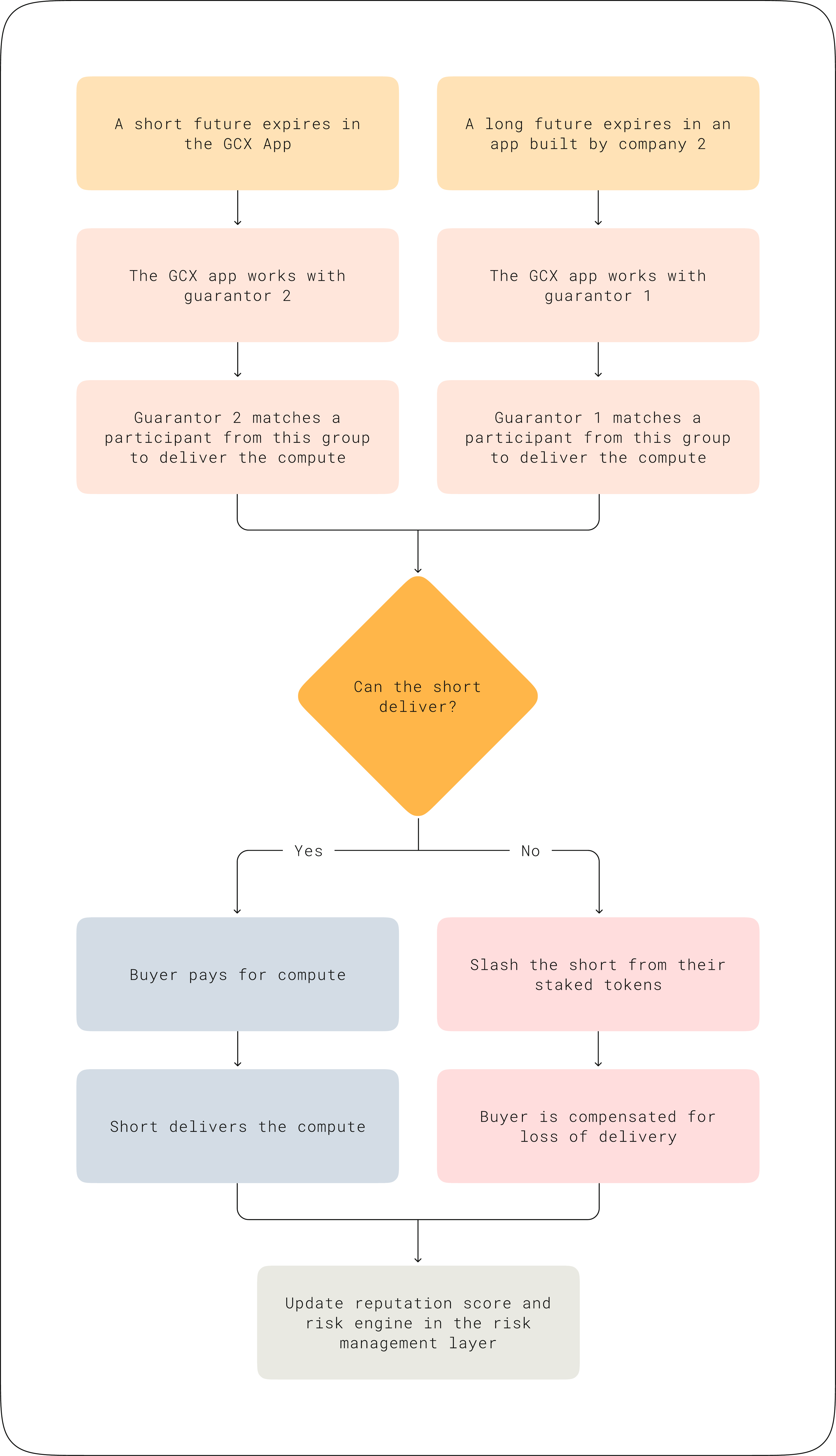}
    \caption{Workflow of the GCX App---Fig.~(\ref{fig:gcx_diagram})---for Short and Long Futures: This process starts with the expiration of a future contract, either short in the GCX App or long in a company-built app. Guarantors match participants to deliver the compute resources. The short participant is required to deliver to the long. If the short can deliver, the buyer pays, and the compute is delivered. If the short fails to deliver, their staked tokens are slashed, and the buyer is compensated. The reputation score and risk engine are updated in the risk management layer accordingly.}
    \label{fig:gcx_participation_selection}
\end{figure}

\subsubsection*{Contract Execution}

When a market participant initiates a trade through the app layer, their guarantor acts as a gatekeeper. Before submitting the trade to the GCX matching engine, the guarantor ensures that the participant has sufficient collateral to cover the associated risk.

\subsubsection*{Compute Delivery Verification}

The selected participant attempts to deliver the compute resources. Successful delivery results in the compute being provided to the buyer, while failure to deliver results in the slashing of the participant’s staked tokens.

\subsubsection*{Exercising an Option and Delivery}

Exercising an option would typically result in the delivery of its underlying futures contract and therefore would not trigger immediate delivery of compute resources. For OTC (over-the-counter) and spot trades, these are settled directly by the counter-parties involved. The primary scenario for delivery to occur is when a futures contract reaches expiration.

\subsubsection*{Futures Contract Expiration and Delivery}

We envision the delivery of compute hours primarily occurring as a result of a GCX futures contract coming to expiration, please see Fig.~(\ref{fig:gcx_participation_selection}).
We take inspiration from existing commodities markets' delivery procedures. For example, in the CME WTI Crude Oil\footnote{These are futures contracts for West Texas Intermediate (WTI) crude oil that are traded on the CME Group’s commodity exchange, which includes the New York Mercantile Exchange (NYMEX) \cite{nymex}. These contracts are used by investors and companies to hedge against price fluctuations in the oil market or to speculate on future price movements.} contract \cite{cme_wti_crude_oil}, buyers can choose from several delivery methods including interfacility transfer, in-facility transfer, or transfer of title \cite{CME200Rulebook}. Similarly, for compute hours, buyers may be offered choices under contract terms specifying different forms of delivery.

Since each guarantor is responsible for its open, deliverable positions at the time of expiration, it will be the guarantor's responsibility to guarantee the delivery of compute hours specified in short futures positions open at expiration. As futures approach expiration, a guarantor may require the holder of short futures positions to verify that they can deliver the contracted compute hours. If they cannot, such short positions can be liquidated to ensure market stability and integrity.

\subsubsection{Verification of Compute and Capacity}

Verification of compute delivery can be automatically ensured using verification protocols similar to those implemented by Truebit \cite{teutsch2017truebit}. Truebit's technology works by having a verifier challenge the computations performed by a solver. If a solver’s computation is challenged and found incorrect, the solver is penalized, ensuring the integrity and correctness of the computations. This mechanism not only ensures that the computation was performed but also verifies its accuracy without the need for all nodes in the network to redo the work. This approach significantly increases the efficiency and scalability of blockchain-based compute platforms.

In addition to verifying that compute tasks were correctly performed, it is also crucial to verify the capacity and availability of compute resources. This means ensuring that the source of compute is indeed available and capable of performing the required tasks. Verification of capacity can be achieved through regular audits and performance checks integrated into the smart contracts. These checks ensure that providers have the necessary resources available and are not oversubscribing their compute capacity. This dual verification process of both compute and capacity enhances the reliability and trustworthiness of the GCX platform.

\subsection{The GCX Utility Token\label{sec:token}}

Over time, one can imagine the GCX evolving into a sovereign chain or rollup, using GCX tokens as the underlying/core token. This would further enhance the platform's efficiency, scalability, and security, providing a robust foundation for global compute trading.

The GCX utility token serves as the backbone of the Global Compute Exchange (GCX) at the Blockchain Layer, offering a versatile and robust utility for our users.
Designed as a utility token, GCX provides access to various functionalities and benefits within our ecosystem, ensuring efficient and effective economic incentives. This utility token is crucial for accessing the exchange's preferred trading rates, collateral for trading activities, and participating in the future app chain.

\subsection*{Membership and Trading Benefits}

GCX tokens can be purchased by users to acquire member seats on the GCX exchange at the Exchange Layer, similar to how seats are bought on traditional exchanges like the NYSE. These seats grant users preferred trading rates and other exclusive privileges, making GCX an essential asset for active traders.

\subsection*{Collateral and Staking}

In addition to membership benefits, GCX tokens are used as collateral for short selling and margin trading. This feature allows traders to leverage their positions while maintaining economic stability within the exchange. Tokens used as collateral are staked, ensuring that operators deliver the required compute resources efficiently. If an operator fails to deliver as promised, their staked tokens can be slashed, providing a built-in mechanism for maintaining accountability and performance \cite{a16zcrypto_slashing}. This aligns with our commitment to creating a transparent and reliable marketplace for computational resources.

\subsection*{Transition to an App Chain}

As we evolve, GCX will transition to becoming the core token of our own app chain. This transition will enhance the token's utility, making it integral to the operation and governance of the new blockchain layer. Users will continue to benefit from the token's utility, with expanded functionalities and opportunities within the app chain ecosystem.

\subsection*{Burn Mechanism and Token Circulation}

GCX tokens have a built-in burn mechanism, ensuring a controlled supply and increasing token value over time. Tokens can be slashed and burned if operators fail to meet performance standards, effectively removing them from circulation.
Additionally, we will introduce new rules for burning and issuance to further stabilize and enhance the GCX ecosystem:

\begin{itemize}
    \item \textbf{Transaction Burn:} A small percentage of tokens will be burned from every transaction on the GCX exchange, reducing the total supply gradually.
    \item \textbf{Performance-Based Burn:} Operators with poor performance metrics may face additional token burns, incentivizing optimal service delivery.
    \item \textbf{Issuance Cap:} New GCX tokens will only be issued based on network growth and demand, maintaining a balanced and sustainable token economy.
    \item \textbf{Operator Rewards:} Operators within the GCX ecosystem may also be rewarded with GCX tokens based on their holdings and performance. This reward system encourages operators to maintain high standards and contribute positively to the network's growth and efficiency.
\end{itemize}

The GCX utility token is a crucial element of our ecosystem, driving membership, trading benefits, collateral management, and operator accountability. With its transition to an app chain and innovative burn mechanisms, our token is poised to become a cornerstone of our growing platform, ensuring long-term value and stability for all users.
This approach not only aligns with current trends in digital transformation but also positions GCX at the forefront of the computational resource market’s evolution.

\subsection*{Conclusion}

The GCX platform provides a versatile platform that caters to various stakeholders in the compute market, enabling them to hedge against price volatility, secure future prices, and generate additional yield. As illustrated in Fig.~(\ref{fig:gcx_diagram}), the GCX allows different actors to leverage its functionalities to their advantage.

Alice, an AI startup founder, uses futures contracts to lock in the price of compute, safeguarding against future price increases \cite{kolb2016futures}. Bob, anticipating a market correction, buys put options to secure a price ceiling, thereby minimizing potential losses \cite{hull2018options}. Carol, a data center operator, sells both call and put options to collect premiums, generating additional yield. By utilizing the GCX, these actors can effectively manage their risks and optimize their financial strategies in the dynamic compute market \cite{passarelli2012trading}.

In all cases, the GCX enables various expressions of market sentiment, allowing participants to profit and protect against the evolution of compute prices, whether they anticipate an increase, decrease, or sideways movement. This flexibility ensures that the GCX can benefit a diverse range of market participants, each with their unique perspectives and strategies.
\section{Contract for Trading Compute Hours}

Contracts between buyers and sellers in a marketplace for computational resources need to be detailed and comprehensive to ensure fairness and stability. Quality specifications should include parameters such as CPU/GPU speed, RAM, and storage capacity. Pricing terms must reflect current market conditions and be adjustable based on supply and demand dynamics. Dispute resolution mechanisms, such as arbitration, should be clearly outlined to handle any conflicts that arise. Leveraging blockchain technology can further enhance contract management by ensuring transparency and immutability \cite{ottina2023, blockchainLegal2019}.
As AI technologies evolve, regulatory frameworks such as the EU AI Act have been developed aiming to ensure ethical and responsible use \cite{euai2023}. These regulations necessitate robust systems for managing and verifying compute resources to ensure compliance.

We will structure the contracts drawing parallels from established practices in power and other commodity markets\footnote{We plan to finalize the contract in consultation with legal counsel, ensuring full compliance with all applicable laws and regulations.} and we can leverage blockchain technology to standardize and track these contracts while maintaining privacy and compliance.

Blockchain technology and smart contracts are already transforming various sectors, proving their utility beyond mere conceptual applications. For instance, in real estate, companies like Propy \cite{propy2019} are facilitating property transactions where the entire buying process, including title transfers, is executed through smart contracts, reducing the transaction time and eliminating traditional intermediaries. In the art and music industry, Glair utilizes blockchain to manage copyrights \cite{fardian2024blockchain, ujoMusic2017}.

In the realm of supply chain management, Everledger leverages blockchain to provide immutable histories of high-value items like diamonds, ensuring the authenticity and legality of the supply chain which aids in compliance with trade regulations \cite{everledger2020}.

Additionally, legal document management systems are beginning to adopt blockchain to secure digital signatures and notarizations that stand up to legal scrutiny and reduce fraudulent activities \cite{blockchainLegal2019, ganne2018}.

The implementation of smart contracts on the GCX platform to manage trading contracts for computational resources is a natural extension of these proven applications. By automating contract execution, ensuring transaction integrity, and providing transparent compliance tracking, smart contracts can offer a robust and efficient framework for trading computational power. This approach not only mitigates risks and lowers operational costs but also aligns with current trends in digital transformation, positioning GCX at the forefront of the computational resource market’s evolution.

Smart contracts on a blockchain can be designed using standardized templates that encapsulate common terms and conditions needed for compute resource trading. By using blockchain, all parties access the same contract versions, eliminating discrepancies and ensuring that transactions are executed based on mutually agreed terms.

Smart contracts can automatically execute transactions based on coded conditions, such as verifying completion and Quality of Service (QoS) adherence before releasing payments. Automation reduces administrative burdens and accelerates processes by eliminating manual steps in contract management and payments.

Every transaction and amendment is recorded on the blockchain, providing a transparent and unchangeable history. Enhanced transparency ensures easy auditability and trust, as all actions are traceable and verifiable.

Smart contracts can include provisions for updates or upgrades as market conditions, regulations, or technology evolves. Updates can be managed through decentralized voting among stakeholders to ensure democratic and agreeable changes.

Blockchain's cryptographic nature makes contracts secure from unauthorized alterations. The immutable and transparent nature of blockchain simplifies the resolution of disputes.

Incorporating blockchain and smart contracts in managing compute resources can significantly enhance efficiency, trust, and adaptability, creating a robust digital trading platform.

\section{The Compute Marketplace}

The GCX trading platform revolutionizes the way compute hour contracts are traded, providing a seamless, secure, and efficient marketplace for both buyers and sellers.
By streamlining the process from request to delivery, we aim to fully ensure that computational resources are utilized effectively, supporting a wide range of industries and applications. Additionally, the GCX's modular approach allows for flexibility and customization, enabling various applications to be built on top of the platform. This means that marketplaces can choose to integrate only specific aspects of the GCX, such as options, futures, perpetual contracts, or any combination thereof. This flexibility ensures that the GCX can meet the diverse needs of different market participants and use cases. In later sections, we seek to illustrate the GCX trading process through detailed examples that encompasses the experiences of both buyers and sellers.
Please consult, e.g., \cite{kolb2016futures, hull2018options, passarelli2012trading} for the underlying derivatives trading theory.

The GCX is designed to accommodate a wide array of compute needs and resource offerings. It integrates advanced matchmaking algorithms with a user-friendly interface to facilitate transactions between buyers and sellers efficiently. The platform supports a variety of contract types, including spot and futures contracts for compute hours, each with customizable terms to suit the specific requirements of users.

Prices for compute hour contracts are determined based on supply and demand dynamics, ensuring fair pricing that reflects current market conditions. All transactions are secured with blockchain technology, providing transparency and integrity to every contract.
The platform's algorithms automatically match sellers' offerings with buyers' needs, optimizing resource allocation and utilization.

\subsection{Spot Trading of Compute}

Spot trading refers to the purchase and sale of commodities for immediate delivery and payment. In the context of compute resources, spot trading involves buying and selling compute hours that can be utilized immediately upon transaction completion. This type of trading is essential for industries and applications that require on-demand computational power without the need for long-term contracts or future deliveries.

While the GCX exchange itself is not technically necessary for spot trading due to the lack of settlement requirements, the GCX app--and, over time, other companies/projects can build their own apps on top of our stack via API and SDK access--serves as a critical marketplace where buyers and sellers can meet, negotiate, and finalize deals. Thus, more generally speaking, the App Layer handles this, see Fig.~(\ref{fig:gcx_diagram}) for the various Layers and how they interact.
The primary objective of the GCX app is to standardize the way buyers and sellers find each other and make transactions, thus facilitating a seamless and efficient marketplace.

The GCX app enables spot trading of compute resources by:

\begin{itemize}
    \item \textbf{Standardizing Contracts}: By standardizing contract terms as much as possible, the GCX app simplifies the negotiation process for counterparties, making it easier for them to agree on deals. This standardization reduces complexity and enhances the efficiency of the marketplace.
    \item \textbf{Facilitating Connections}: The app provides a platform where buyers and sellers can easily find each other and negotiate terms. It acts as a matchmaking service, ensuring that computational needs are met with available resources.
    \item \textbf{Incentivizing Participation}: To encourage trading activity, the GCX app rewards participants with tokens for their trades. These incentives promote engagement and contribute to the success of the platform.
\end{itemize}

By focusing on pure over-the-counter (OTC) spot trading in the beginning stages of the GCX app we eliminate the need for clearing pools or settlement processes on day one.
Instead, the GCX app provides a streamlined environment for buyers and sellers to find each other and negotiate deals. This approach allows the platform to learn from real-world trading activity and adapt to the needs of its users.

\subsection{Derivatives Trading of Compute}

Derivatives trading involves financial contracts whose value is derived from an underlying asset—in this case, compute hours. Common derivatives include futures, options, and perpetual contracts (perps). These instruments allow market participants to hedge against price fluctuations, speculate on future price movements, and manage risk more effectively.

\begin{itemize}
    \item \textbf{Futures Contracts}: Agreements to buy or sell compute hours at a predetermined price on a specified future date. These contracts help participants lock in prices and secure compute resources for future use.
    \item \textbf{Options Contracts}: Provide the right, but not the obligation, to buy or sell compute hours at a predetermined price before a specified expiration date. Options offer flexibility and risk management strategies for market participants.
    \item \textbf{Perpetual Contracts (Perps)}: Similar to futures but without an expiry date. They enable continuous trading of compute hours and are settled periodically based on market conditions. This can be enabled via the Blockchain Layer, see Fig.~(\ref{fig:gcx_diagram}).
\end{itemize}

In contrast to spot trading, which involves immediate delivery, derivatives trading allows participants to plan and strategize for future computational needs and costs. Spot trading is essential for derivatives because it provides the actual compute hours that underlie the derivatives contracts, ensuring that the market has the necessary liquidity and resource availability.

The GCX exchange is crucial for derivatives trading as it handles the settlement of these contracts. By evolving from the insights gained through the GCX app, the GCX exchange can develop into a robust platform capable of managing the complexities of derivatives trading.

The GCX platform supports derivatives trading by offering:

\begin{itemize}
    \item \textbf{Comprehensive Contract Types}: A variety of derivatives contracts, including futures, options, and perps, each customizable to meet the specific needs of buyers and sellers.
    \item \textbf{Advanced Trading Tools}: Features such as real-time market data, analytical tools, and risk management options to help participants make informed trading decisions.
    \item \textbf{Onchain Trading}: The integration of blockchain technology allows for secure, transparent, and efficient onchain trading of derivatives, ensuring the integrity of every contract.
\end{itemize}

By enabling both spot and derivatives trading, the GCX platform overall provides a complete ecosystem for trading compute resources. Spot trading ensures immediate access to compute hours, while derivatives trading offers tools for future planning and risk management. Together, these capabilities support a wide range of market participants, from those needing immediate computational power to those seeking to hedge against future price fluctuations or invest in the compute market.


In the following sections, we will explore several examples of both spot and derivatives trading, illustrating how the GCX platform’s commodification of compute can revolutionize access to computational resources, resulting in significant impacts across various industries:

\subsection{Example 1: Hedging Strategies for Compute Resource Volatility}

Let's consider a specific hedging example where Alice wants to mitigate compute resource costs for her startup \cite{kolb2016futures}:

\subsubsection*{Alice's Scenario: Buy-side Hedge}

Alice is the founder of an AI-focused startup seeking venture funding. She anticipates needing 500 TFLOP Hours per month starting in six months, with an ongoing need of 1,000 TFLOP Hours per month after six months of training. To mitigate the risk of higher than expected compute costs, Alice can use the GCX platform to hedge against price volatility \cite{jindal2024compute}.

\subsubsection*{Option 1: Futures Contracts}
Alice can purchase futures on the GCX marketplace to lock in her expected compute needs. This allows her to:
\begin{itemize}
  \item Purchase more or sell excess compute power if her needs change.
  \item Roll her existing futures forward or backward to adjust the timing of delivery.
  \item Take delivery of the compute hours at the contract price or sell the futures to lock in gains if prices rise.
\end{itemize}

\subsubsection*{Option 2: Options Strategies}
Alice can use various options strategies to protect against changes in her compute needs:
\begin{itemize}
  \item Buy Call options to protect against increased prices for additional compute hours.
  \item Buy Put options to protect against needing less compute and falling prices.
  \item Use a combination of Call and Put options (Straddle or Strangle) to protect against both upside and downside changes in compute needs.
\end{itemize}

\subsubsection*{Advantages}
\begin{itemize}
  \item Alice can reduce her exposure to changing compute prices and requirements.
  \item She can focus on her business without worrying about price volatility.
  \item Venture capitalists are more comfortable investing due to reduced risk.
\end{itemize}

\subsubsection*{Disadvantages}
\begin{itemize}
  \item Trading on an exchange may subject Alice to maintenance margin requirements.
  \item Her needs may not match exactly with standard futures contracts.
\end{itemize}

\subsubsection*{Sell-side Hedge: Bob's Scenario}

Bob is building new datacenters for his cloud-based compute business. He needs to lock in prices for future compute sales to secure financing and mitigate the risk of falling prices \cite{smith2018resource}.

\subsubsection*{Option 1: Sell Futures Contracts}
Bob can sell long-term futures on the GCX marketplace to lock in prices. As contracts near expiration, they can be rolled into shorter-term contracts.

\subsubsection*{Option 2: Buy Put Options}
Bob can buy put options on his futures contracts to protect against downside price risk while allowing for upside gains.

\subsubsection*{Advantages}
\begin{itemize}
  \item Bob can confidently build new datacenters knowing he has locked in profitable pricing.
  \item Lenders are more likely to finance his venture due to reduced risk.
\end{itemize}

\subsubsection*{Disadvantages}
\begin{itemize}
  \item Selling futures or call options may lead to large margin calls if compute prices rise rapidly.
  \item Standardized contracts may not match Bob’s specific needs.
\end{itemize}

As mentioned, another advantage for both Alice’s and Bob’s businesses is that their investors experience reduced price volatility and more predictable financial outcomes, making these ventures more attractive and secure.


\subsection{Example 2: Optimizing Yield for Datacenter Operators}

Carol owns a large datacenter and is currently profitable based on current compute pricing. A significant cost for Carol to run her datacenter is the cost of power, so she can choose to turn off her datacenter if power costs outweigh the compute revenues. Carol would like to take advantage of the optionality embedded in being able to power up and down her datacenter. She is profitable selling at current compute prices and would like to earn additional yield to generate steady income and smooth out her long-term revenues \cite{passarelli2012trading}.

\subsubsection*{Option 1: Selling Call Options}
Carol can sell strips of call options, which will generate a cash premium when sold. This strategy is most advantageous when volatility in compute prices is high since option prices are highly correlated with expected volatility in the underlying asset’s price. Selling options at times of high volatility will both take advantage of the high premium to generate additional yield, while at the same time reducing the variability in Carol’s future cash flows. Carol can sell call options with strikes set to the current futures price of compute, or she can opt to sell options with strikes that are higher than the current price of compute.

The choice of strike price is dependent on the option volatility smile (skew). If the volatility of higher strike call options is greater than the at-the-money strikes, then Carol can take advantage of this condition to gain additional premium by selling these options. This situation can develop when there is high uncertainty in the markets and fear of prices spiking higher. If the volatility of higher strike call options is lower than the at-the-money strikes, then Carol may wish to sell call options closer to at-the-money. This is the typical situation in many markets due to the natural tendency of producers to hedge.

\subsubsection*{Option 2: Selling Both Puts and Calls}
To provide additional yield, Carol can take on a riskier position by selling puts in addition to calls. This strategy would be more advantageous in a market where option put prices are trading at a significant volatility premium relative to call prices, often the case in a producer-hedger dominated market in times of lower volatility. By selling puts, Carol may be required to purchase compute at a price that, while lower than today’s prices, may be significantly lower than the strike price of the put option she sold, thus resulting in significant losses.

However, Carol has the option to turn off her datacenter at these times, so while she will end up losing money on the options if prices drop, she can shut down her own production, putting her in a position as if she were simply unhedged on downside price moves from the start. This strategy carries risks but can be very advantageous when there is a strong put skew in the market. Carol can offset some of this risk by buying an additional put option at a strike below the put option she sold. This will reduce her yield but can return her to a fully hedged position, allowing her to take advantage of yield and smooth out her returns.

\paragraph{Advantages:}
\begin{itemize}
    \item Selling options generates a premium that provides additional cash flow up front, offering more working capital.
    \item Selling call options helps smooth out revenue streams, beneficial for generating more consistent earnings over time.
    \item Selling put options allows Carol to generate additional yield from the optionality embedded in the ability to turn off her production at any time.
\end{itemize}

\paragraph{Disadvantages:}
\begin{itemize}
    \item Selling calls may result in missed opportunities to sell compute at very high prices during extreme demand spikes.
    \item Selling puts may result in large losses if the market price of compute drops significantly, effectively losing the optionality of turning off compute and leaving the downside unhedged.
\end{itemize}

In summary, the GCX marketplace represents a significant leap forward in the trading of computational resources. By leveraging blockchain technology and smart contracts, we ensure that all transactions are secure, transparent, and efficient.
This innovative approach facilitates dynamic and fair pricing but also enhances trust and reliability of the platform.
As demonstrated through our detailed example, the GCX platform is poised to support a wide array of compute needs across diverse industries, driving the commodification of compute resources and unlocking new possibilities for growth and innovation.


\subsection{Example 3: Enabling Financial Trading on the Global Compute Exchange}

Amazon Web Services (AWS) \cite{awsdocs} is a leading supplier of cloud computing resources, while we imagine that Citadel \cite{citadel}, a major hedge fund, engages in financial trading of compute resources on the GCX. Citadel makes strategic bets on the future direction of compute prices, requiring large quantities of compute resources to support its trading activities. To facilitate these trades, Citadel purchases substantial compute resources from AWS \cite{hull2018options}.

AWS provides a vast array of compute resources on the GCX platform. Citadel, operating as a financial trader, needs to acquire large quantities of these resources to execute its trading strategies. By leveraging the GCX, Citadel can efficiently buy and sell compute resources, similar to how it would trade commodities like oil.

\subsubsection*{Option 1: Long-term Contracts with AWS}
Citadel can enter into long-term contracts with AWS on the GCX platform. This allows Citadel to:
\begin{itemize}
    \item Secure compute resources at a fixed price, protecting against future price increases and enabling predictable financial planning.
    \item Ensure a consistent supply of compute resources, essential for maintaining trading volumes and liquidity in the compute market.
\end{itemize}

\subsubsection*{Option 2: Spot Market Purchases}
Citadel can also utilize the spot market on the GCX platform to buy compute resources from AWS as needed. This strategy offers:
\begin{itemize}
    \item Flexibility to purchase compute power based on real-time demand, potentially benefiting from lower prices during periods of low demand.
    \item The ability to dynamically adjust resource levels to align with market conditions and trading activity.
\end{itemize}

\subsubsection*{Advantages}
\begin{itemize}
    \item GCX provides a transparent and efficient marketplace for trading compute resources, ensuring fair pricing and reliable transactions.
    \item Long-term contracts with AWS offer Citadel price stability and secure supply, crucial for executing large trades.
    \item The spot market allows Citadel to optimize costs by purchasing compute resources at favorable prices during ``compute off-peak" times.
\end{itemize}

\subsubsection*{Disadvantages}
\begin{itemize}
    \item Long-term contracts may limit Citadel's ability to take advantage of potential future price drops in the spot market.
    \item Relying solely on the spot market can expose Citadel to price volatility and potential shortages during high demand periods.
\end{itemize}

Through the GCX, Citadel can efficiently manage its compute resource portfolio by purchasing in bulk from AWS and leveraging these resources for financial trades. This enables Citadel to place large bets on the future direction of compute prices, providing liquidity and depth to the market. The use of smart contracts ensures that transactions are executed automatically and transparently, with clearly defined terms and conditions.

Both AWS and Citadel benefit from the security and compliance features embedded in the GCX platform. Smart contracts ensure that all transactions are secure, transparent, and auditable, meeting regulatory requirements and protecting sensitive data.

By leveraging the Global Compute Exchange, AWS can supply compute resources to Citadel, enabling high-volume financial trading and liquidity in the compute market. GCX provides a robust and flexible platform for managing compute resource transactions, offering stability, flexibility, and security in the dynamic compute market.

\subsection{Example 4: Enhancing Financial Analytics for Fintech Startups}

Fintech AI modeling startup Nearest helps wealth managers better track portfolios, identify trends, and mitigate risks using sophisticated analytics on historical timeseries data. However, the small data science team struggled with long lead times for accessing GPU servers for model training experiments on Azure, costing tens of thousands of dollars despite modest needs. By leveraging the GCX compute application's (built on top of the GCX exchange\footnote{The ``GCX compute application" is in the GCX App Layer whereas the ``GCX Exchange" is in the Exchange Layer.}) auto-scaling pools with cloud credits, they could run iterations much faster without the hassles of managing underlying infrastructure.

Nearest, an AI-focused fintech startup, requires significant computational resources for running complex financial models. Traditional cloud services like Azure were proving too costly and time-consuming, impeding their ability to innovate and deliver timely insights to clients.

\subsubsection*{Advantages}
\begin{itemize}
\item Nearest can speed up their model training processes, delivering insights to clients faster.
\item Reduced infrastructure management burden allows the team to focus on enhancing their analytics capabilities.
\item Cost savings from cloud credits and dynamic scaling improve financial efficiency and flexibility.
\end{itemize}

\subsubsection*{Disadvantages}
\begin{itemize}
\item Relying on cloud credits and dynamic scaling may lead to unpredictable costs if not managed carefully.
\item The performance of auto-scaling pools can vary based on market demand and availability.
\end{itemize}

In summary, the GCX marketplace enables fintech startups like Nearest to overcome the high costs and inefficiencies associated with traditional cloud services. By leveraging auto-scaling pools and cloud credits, Nearest can enhance their financial analytics capabilities, providing better services to their clients while maintaining cost-effectiveness and operational efficiency.

\subsection{Example 5: On-Demand Rendering for Media Production}

Animation studio Pixar needed on-demand rendering capacity for a new animation film without major upfront capital costs before release. By leveraging the GCX platform, which links low-priority capacity across render farms, Pixar saved 30\% in costs with a faster turnaround for the CGI-heavy project by paying only for the exact hours consumed. The auto-scaled capacity also reduced risks of output delays or resource saturation issues.

\subsubsection*{Pixar’s Scenario: Utilizing the GCX Application}

Pixar, a leader in animation production, requires extensive computational resources for rendering high-quality animation films. Traditional methods involving significant upfront capital investments in rendering farms were not feasible for the tight production timelines and budget constraints of their new project.

\subsubsection*{Leveraging Low-Priority Capacity}
Pixar can use the GCX marketplace to access low-priority capacity across render farms. This allows them to:
\begin{itemize}
\item Utilize idle compute resources for rendering tasks, significantly reducing costs.
\item Scale resources dynamically to meet the demands of the project, ensuring timely completion.
\item Pay only for the exact hours consumed, optimizing their budget and avoiding unnecessary expenditures.
\end{itemize}

\subsubsection*{Advantages}
\begin{itemize}
\item Pixar achieves significant cost savings by utilizing low-priority capacity and paying only for consumed resources.
\item The auto-scaling feature ensures timely project completion, maintaining high-quality output.
\item Reduced infrastructure management allows Pixar to allocate more resources to creative development.
\end{itemize}

\subsubsection*{Disadvantages}
\begin{itemize}
\item Depending on low-priority capacity may lead to variability in resource availability during peak times.
\end{itemize}

In summary, the GCX marketplace provides a robust solution for media production companies like Pixar, enabling them to access on-demand rendering capacity without the burden of significant upfront capital costs. By leveraging low-priority capacity, Pixar can deliver high-quality animation projects efficiently and cost-effectively.

\section{Conclusion}

We have introduced the Global Compute Exchange (GCX) platform, a groundbreaking solution for the commodification and efficient trading of computational resources. By standardizing compute units through Compute Hours (CH) we have created a robust framework that ensures fair, transparent, and efficient trading. This approach allows for a seamless comparison and utilization of diverse computing resources, much like the established practices in electricity markets.
The GCX helps to address critical inefficiencies in the current compute landscape, such as underutilization and price volatility, ensuring equitable access to computational power, stimulating innovation, and supporting diverse user needs on a global scale.

The GCX platform’s architecture is divided into several layers, each with specific functions to ensure a robust and efficient operation:

\begin{itemize}
    \item \textbf{Market Layer}: Comprises participants such as hedgers, traders, short sellers, and market makers who interact with the GCX to trade compute resources.
    \item \textbf{App Layer}: Consists of applications developed by various companies, including the GCX app, which provides user interfaces and frontends for interacting with the GCX platform. This is where spot trading of compute takes place natively. Derivatives trading also happens here, but in this case it settles on the Exchange Layer below.
    \item \textbf{Clearing Layer}: Facilitates smooth and secure trade settlements with guarantors ensuring the delivery of compute resources and managing collateral.
    \item \textbf{Risk Management Layer}: Monitors and manages risks associated with trading on the GCX, including APIs, SDKs, and a risk engine for monitoring and analytics.
    \item \textbf{Exchange Layer (Offchain)}: Manages the off-chain operations of the GCX, including matching buy and sell orders and executing trades.
    \item \textbf{Blockchain Layer (Onchain)}: Manages onchain operations, leveraging blockchain technology for transparency and security, including smart contracts, token collateral, and staking mechanisms.
\end{itemize}

The GCX App Layer facilitates spot trading by standardizing contract terms, enabling buyers and sellers to efficiently negotiate deals and execute trades. This approach not only enhances market efficiency but also provides valuable insights into market dynamics, helping to inform the development of future products and services. Incentivizing participants with tokens further promotes engagement and contributes to the platform’s success.

Derivatives trading in the GCX App Layer settled in the Exchange Layer introduces sophisticated financial instruments such as futures, options, and perpetual contracts. These tools allow participants to hedge against price fluctuations, speculate on future price movements, and manage risk effectively. While the trading of derivatives takes place on the GCX app, the settlement of these contracts is managed by the GCX exchange. The integration of blockchain technology ensures secure, transparent, and efficient onchain trading, maintaining the integrity of every contract.

Through spot and derivatives trading, GCX provides a complete ecosystem for trading compute resources, supporting a wide range of market participants. This dual capability ensures immediate access to compute hours for urgent needs and offers strategic tools for future planning and risk management.

The GCX platform’s potential to revolutionize access to computational resources is profound. 
By turning compute hours into a tradable commodity, GCX optimizes resource utilization, stabilizes pricing, and democratizes access to computational power. This democratization is crucial for enabling innovation across various sectors, from startups to large enterprises, fostering a more inclusive and dynamic technological landscape.

\section{Glossary}
\begin{itemize}

\item \textbf{CPU (Central Processing Unit)}: The primary component of a computer that performs most of the processing inside a computer. It executes instructions from programs, performing basic arithmetic, logic, control, and input/output (I/O) operations. CPUs are characterized by their versatility and are designed to handle a wide variety of tasks in computing systems.

\item \textbf{GPU (Graphics Processing Unit)}: A specialized processor designed to accelerate graphics rendering. GPUs can process many parallel tasks more efficiently than CPUs, making them ideal for handling complex computations in graphics rendering, machine learning, and scientific simulations. They are particularly known for their ability to perform floating-point calculations required in rendering 3D graphics and AI.

\item \textbf{GPGPU (General-Purpose computing on Graphics Processing Units)}: GPGPU is the utilization of a GPU, which is typically used for rendering graphics, to perform computation traditionally handled by the CPU. This technique leverages the parallel processing capabilities of GPUs to accelerate a wide range of applications beyond graphics, including scientific simulations, machine learning, data analysis, and more.

\item \textbf{AI (Artificial Intelligence)}: The simulation of human intelligence processes by machines, especially computer systems. AI involves the development of algorithms and software to perform tasks that typically require human intelligence, such as visual perception, speech recognition, decision-making, and language translation. Key areas of AI include machine learning, where systems learn and improve from experience, and deep learning, which involves neural networks with many layers.

\item \textbf{Compute Hour (CH)}: A unit of measure representing the computational effort equivalent to one hour of processing on a reference system with a specified performance, typically measured in teraFLOPS (trillions of floating-point operations per second). Example: If a system performs 2 teraFLOPS, it will generate 2 Compute Hours per hour of operation.

\item \textbf{Epochs}: An epoch refers to one complete pass through the entire training dataset. During each epoch, the model is trained on all the training data once. Training typically involves multiple epochs, and the number of epochs is a crucial hyperparameter that impacts the model’s learning and generalization abilities. Too few epochs can lead to underfitting, where the model does not learn adequately from the training data, while too many epochs can cause overfitting, where the model learns noise and performs poorly on new data.

\item \textbf{Batch Size}: The batch size is the number of training samples processed before the model’s internal parameters are updated. Smaller batch sizes can make training noisier but help models generalize better, whereas larger batch sizes can make the training process faster but may lead to overfitting. The choice of batch size affects the stability and speed of the training process and is another critical hyperparameter that needs careful tuning.

\item \textbf{Iterations}: An iteration refers to one update of the model’s parameters and is completed once a single batch of data has been processed. The number of iterations per epoch is determined by the size of the dataset divided by the batch size. For instance, if the dataset contains 1,000 samples and the batch size is 100, it will take 10 iterations to complete one epoch.

\item \textbf{Smart Contract}: Self-executing contracts with the terms of the agreement directly written into code. These contracts automatically enforce and execute the terms when predefined conditions are met, running on a blockchain to ensure transparency and immutability. Example: A smart contract on the GCX platform could automatically transfer payment to a compute provider once the specified compute hours are delivered and verified.

\item \textbf{Blockchain}: A decentralized digital ledger that records transactions across many computers in such a way that the registered transactions cannot be altered retroactively. This technology ensures transparency, security, and trust in the data recorded. Example: Blockchain technology underpins the GCX platform, ensuring that all transactions are secure and transparent.

\item \textbf{Token Staking}: The process of locking tokens in a blockchain-based system to support the network's operations, such as transaction validation, and in return, earning rewards. In the context of GCX, operators stake tokens as a commitment to provide computational resources. Example: Operators on GCX stake tokens to join a compute pool, which are slashed if they fail to deliver the promised compute hours.

\item \textbf{Decentralized Governance}: A governance structure where decision-making is distributed among all stakeholders rather than being centralized in a single entity. This approach often involves voting mechanisms to make decisions on protocol updates, policies, and other governance matters. Example: GCX may use decentralized governance to allow token holders to vote on changes to smart contracts or platform policies.

\item \textbf{GCX (Global Compute Exchange)}: A platform facilitating a marketplace designed to standardize and commoditize computational resources. GCX facilitates the buying and selling of compute resources by providing a transparent and efficient trading system. It leverages standardized benchmarks to measure and trade compute units, ensuring fair and consistent evaluation of computational power across various providers.

\item \textbf{Dynamic Pricing}: A pricing strategy where the price of a commodity fluctuates based on real-time supply and demand conditions. In the GCX marketplace, dynamic pricing ensures fair market value for compute hours. Example: The price of compute hours on GCX might increase during peak demand periods and decrease when demand is low.

\item \textbf{Predictive Analytics}: The use of statistical algorithms and machine learning techniques to analyze historical data and make predictions about future events. In GCX, predictive analytics help anticipate compute resource needs and preempt potential failures. Example: GCX uses predictive analytics to forecast demand for compute hours and ensure resources are allocated efficiently.

\item \textbf{FLOP:} A FLOP, or Floating Point Operation (and we take TFLOPs to be the plural form, namely, Floating Point Operations), is a measure of computational performance, especially crucial in tasks involving real-number calculations common in AI and machine learning algorithms. It quantifies the number of calculations a system can perform per second, with higher units like teraFLOPS (TFLOPs) representing one trillion FLOPs and petaFLOPS (PFLOPs) representing one quadrillion FLOPs. Supercomputers performing hundreds of petaFLOPs can handle complex simulations and large-scale data analyses. In AI, models requiring extensive computational power, such as deep learning networks, often measure their demands in teraFLOPs or even petaFLOPs, reflecting the immense resources needed for processing and training.

\item \textbf{FLOP-Hours}: A measure of computational work that represents the number of floating-point operations a system can perform in an hour. It is used to quantify and compare the computational performance of different systems. Example: A system that performs 1 trillion floating-point operations per second (1 TFLOP) over one hour would generate 1 FLOP-Hour.

\item \textbf{PetaFLOP (PFLOP)} A PetaFLOP is a measure of computational speed, representing one quadrillion ($10^{15}$) floating-point operations per second (FLOPS). It is used to quantify the performance of supercomputers and other high-performance computing systems. A supercomputer with a peak performance of 1 PetaFLOP can perform $10^{15}$ floating-point calculations every second.

\item \textbf{FLOPS (Floating Point Operations Per Second)}: A measure of computer performance, especially in fields of scientific calculations that require floating-point calculations. It indicates how many floating-point arithmetic operations a computer can perform in one second. Higher FLOPS values indicate greater computational power.

\item \textbf{Call Option}: A financial contract that gives the buyer the right, but not the obligation, to purchase a specified amount of compute resources at a predetermined price within a specified time period. This is used to hedge against or speculate on price increases.

\item \textbf{Put Option}: A financial contract that gives the buyer the right, but not the obligation, to sell a specified amount of compute resources at a predetermined price within a specified time period. This is used to hedge against or speculate on price decreases.

\item \textbf{Futures Contract}: A standardized legal agreement to buy or sell compute resources at a predetermined price at a specified time in the future. This allows companies to hedge against price volatility and secure stable compute costs.

\item \textbf{Spot Market}: A public financial market in which compute resources are traded for immediate delivery. Prices in the spot market are determined by current supply and demand.

\item \textbf{Spot versus Derivatives:} Spot trading involves immediate delivery of assets, while derivatives trading involves contracts based on future asset delivery and price speculation, such as futures and options.

\item \textbf{OTC (Over-the-Counter)}: OTC refers to the process of trading financial instruments directly between two parties without the supervision of an exchange. These trades occur in a decentralized market where participants trade directly with one another, typically through a network of dealers and brokers.

\item \textbf{Volatility Smile (Skew)}: A pattern in which options with different strike prices exhibit varying levels of implied volatility. Higher strike prices may have different implied volatilities than at-the-money strikes, affecting the pricing and attractiveness of these options.

\item \textbf{Straddle Option}: An options strategy involving the purchase of both a call and put option with the same strike price and expiration date. This is used to profit from large price movements in either direction.

\item \textbf{Strangle Option}: An options strategy involving the purchase of a call and put option with different strike prices but the same expiration date. This is used to profit from large price movements in either direction, with a lower cost than a straddle.

\item \textbf{Dynamic Pricing System}: A mechanism that adjusts the price of compute resources in real-time based on current market supply and demand conditions, ensuring fair and transparent transactions.

\item \textbf{Smart Contract}: A self-executing contract with the terms of the agreement directly written into code. Smart contracts automatically execute and enforce agreements when predetermined conditions are met, ensuring transparency and reducing the need for intermediaries.

\item \textbf{Decentralized Task Queue}: A system that allows users to submit compute tasks to a distributed network, where tasks are dynamically allocated to available resources based on a consensus mechanism.

\item \textbf{Resource Matching Queue}: A decentralized system that matches and allocates compute resources to tasks based on real-time demand and resource availability, optimizing resource utilization.

\item \textbf{Hedging}: The practice of making financial investments to reduce the risk of adverse price movements in an asset. In the context of compute resources, hedging involves using financial instruments like futures and options to manage price volatility.

\item \textbf{Yield Generation}: The process of earning returns on an investment. In the context of compute resources, this involves strategies like selling options to generate additional income and stabilize revenue streams.

\item \textbf{Embedded Optionality}: The inherent flexibility in an asset or operation that allows the owner to make strategic decisions based on changing conditions. For example, the ability to power up or down a datacenter in response to power costs and compute revenues.

\item \textbf{Carbon Footprint}: The total amount of greenhouse gases produced directly or indirectly by human activities. In the context of compute resources, managing and reducing the carbon footprint is crucial for sustainability.

\item \textbf{Compute as a Commodity (CaaC):} A paradigm that treats computing power as a commodity, similar to utilities like electricity and water. CaaC allows organizations to access vast compute resources on-demand and pay only for what they use, without the need for significant upfront investments. This model leverages the scalability and flexibility of cloud computing, enabling businesses to innovate more rapidly and efficiently by aggregating underutilized capacity across multiple providers and geographies.

\item \textbf{Compute as a Service (CaaS):} A cloud computing model that provides computing resources as a service over the internet. CaaS allows businesses to rent processing power, storage, and networking infrastructure on a pay-as-you-go basis. This model simplifies the deployment and management of IT resources, offering scalability, cost efficiency, and flexibility. Organizations can quickly scale their infrastructure to meet changing demands without the need for substantial capital investments in physical hardware.

\item \textbf{Infrastructure as a Service (IaaS):} A cloud computing model that provides virtualized computing resources over the internet. IaaS offers fundamental IT resources such as virtual machines, storage, and networking on a pay-as-you-go basis. It allows businesses to rent and manage these resources without having to purchase and maintain physical hardware. IaaS is highly scalable and flexible, enabling organizations to dynamically adjust their infrastructure to meet changing demands. Examples of IaaS providers include Amazon Web Services (AWS), Microsoft Azure, and Google Cloud Platform (GCP).

\item \textbf{Platform as a Service (PaaS):} A cloud computing model that provides a platform allowing customers to develop, run, and manage applications without dealing with the underlying infrastructure. PaaS offers a complete development and deployment environment, including tools, libraries, services, and infrastructure, to support the entire application lifecycle. This model abstracts much of the complexity of building and maintaining the underlying hardware and software layers, allowing developers to focus on coding and innovation. Examples of PaaS providers include Google App Engine, Microsoft Azure App Services, and Heroku.

\item \textbf{Perps (Perpetual Futures):} A type of derivative contract in financial markets that has no expiration date. Traders can hold positions indefinitely, allowing for continuous speculation on the price of an underlying asset, often with leverage.

\item \textbf{Perpetual Options:} Financial derivatives that grant the holder the right, but not the obligation, to buy or sell an underlying asset at any time without an expiration date, providing flexibility and continuous trading opportunities.

\item \textbf{DeFi (Decentralized Finance):} A financial ecosystem built on blockchain technology that eliminates intermediaries, offering a wide range of financial services such as lending, borrowing, and trading through decentralized protocols and smart contracts. This technology also allows for a seamless integration of Perps and Perpetual Options to be implemented and traded on the GCX.

\item \textbf{Market Layer}: This layer consists of various market participants including hedgers, proprietary traders, short sellers, and market makers who own GCX tokens. These participants engage in trading activities and strategies within the compute resource market.

\item \textbf{App Layer}: This layer comprises the user interfaces and frontend applications, including GCX's app and other companies' apps. It also includes the APIs and SDKs necessary for developing and integrating applications with the Global Compute Exchange.

\item \textbf{Clearing Layer}: The layer responsible for ensuring the verification and delivery of compute resources. It includes guarantors who provide proof of compute capacity and manage the delivery of compute resources, ensuring the integrity and reliability of the transactions.

\item \textbf{Risk Management Layer}: This layer focuses on identifying, assessing, and mitigating risks associated with trading compute resources. It involves risk identification, data collection and monitoring, risk assessment, and the implementation of risk models and mitigation strategies.

\item \textbf{Exchange Layer (Offchain)}: The offchain component of the Global Compute Exchange that handles the trading infrastructure, including the Exchange API and SDK. It supports various marketplaces for spot, derivatives, futures, and options trading.

\item \textbf{Blockchain Layer (Onchain)}: The foundational layer utilizing blockchain technology to ensure secure, transparent, and immutable transactions. It includes a smart contract ecosystem, token interfaces, and collateral mechanisms such as staking, slashing, and insurance pools. This layer integrates with the broader DeFi ecosystem and manages customer and guarantor collateral.

\item \textbf{Offchain}: Refers to processes and transactions that occur outside the blockchain network. In the context of the Global Compute Exchange, offchain activities include the trading infrastructure, APIs, and SDKs that facilitate the exchange of compute resources without directly recording each transaction on the blockchain. This allows (for now) for faster and more scalable operations, with periodic synchronization to the blockchain for settlement and verification.

\item \textbf{Onchain}: Refers to processes and transactions that are executed and recorded directly on the blockchain network. In the context of the Global Compute Exchange, onchain activities include the use of smart contracts, token interfaces, and collateral management. These onchain mechanisms ensure transparency, security, and immutability, as every transaction is permanently recorded on the blockchain, enabling decentralized and trustless interactions.

\item \textbf{Blockchain Network}: A decentralized digital ledger that records transactions across many computers in such a way that the registered transactions cannot be altered retroactively. Each block contains a cryptographic hash of the previous block, a timestamp, and transaction data. Examples include:
    \begin{itemize}
        \item \textbf{Ethereum}: A blockchain network that supports smart contracts and decentralized applications (dApps) \cite{buterin2014ethereum}. It features its native cryptocurrency, Ether (ETH), and enables developers to build and deploy their own applications on the network.
        \item \textbf{Bitcoin}: The first and most well-known blockchain network, primarily used for peer-to-peer transactions with its native cryptocurrency, Bitcoin (BTC) \cite{nakamoto2008bitcoin}. Bitcoin's blockchain focuses on providing a secure and decentralized way to transfer value and store data.
    \end{itemize}

\item \textbf{Message Passing Interface (MPI)}: MPI is a standardized and portable message-passing system widely used in parallel computing. It enables processes to communicate with each other by sending and receiving messages, making it suitable for various parallel computing architectures. MPI is designed to be scalable, high-performing, and flexible, supporting both point-to-point and collective communication. It also provides mechanisms for creating process groups and communicators, facilitating structured communication in large-scale applications.

\item \textbf{DVS (Dynamic Voltage Scaling)}: A power management technique that dynamically adjusts the voltage supplied to a processor based on performance needs. This helps in reducing power consumption and heat generation, thus enhancing energy efficiency, especially in battery-powered devices.

\item \textbf{DVFS (Dynamic Voltage and Frequency Scaling)}: An advanced power management technique that adjusts both the voltage and frequency of a processor in response to workload demands. By scaling down voltage and frequency during low workloads, DVFS optimizes energy consumption and improves overall efficiency in computing systems.

\end{itemize}

\bibliographystyle{unsrt}  
\bibliography{references}  

\begin{thebibliography}{100}

\bibitem{sevilla2022compute}
Jaime Sevilla, Lennart Heim, Anson Ho, Tamay Besiroglu, Marius Hobbhahn, and Pablo Villalobos.
\newblock Compute trends across three eras of machine learning.
\newblock \url{https://arxiv.org/pdf/2202.05924.pdf}, 2022.
\newblock Accessed: 2024-07-02.

\bibitem{Appenzeller2023}
Guido Appenzeller, Matt Bornstein, and Martin Casado.
\newblock Navigating the high cost of ai compute.
\newblock {\em Andreessen Horowitz}, April 2023.

\bibitem{Constine2023}
Josh Constine and Veronica Mercado.
\newblock The ai compute shortage explained by nvidia, crusoe, \& mosaicml.
\newblock {\em SignalFire Blog}, August 2023.

\bibitem{awsdocs}
{Amazon Web Services}.
\newblock {\em Amazon Web Services (AWS) Documentation}, 2024.
\newblock Available at: \url{https://docs.aws.amazon.com/}, Accessed: 2024-06-12.

\bibitem{Dave2023}
Paresh Dave.
\newblock Nvidia chip shortages leave ai startups scrambling for computing power.
\newblock {\em Wired}.
\newblock Accessed: 2024-07-02.

\bibitem{ChatGPT}
OpenAI.
\newblock Chatgpt: Conversational ai model.
\newblock \url{https://www.openai.com/chatgpt}, 2023.
\newblock Accessed: 2024-07-02.

\bibitem{nvidia2024}
NVIDIA.
\newblock Nvidia official website, 2024.
\newblock Available at: \url{https://www.nvidia.com}, Accessed: 2024-06-23.

\bibitem{timofeev2024case}
Paul Timofeev.
\newblock The case for compute depins.
\newblock {\em Shoal Research}, 2024.
\newblock Available at: \url{https://www.shoal.gg/p/the-case-for-compute-depins}, Accessed: 2024-06-12.

\bibitem{jindal2024compute}
Siddharth Jindal.
\newblock Compute is the new oil.
\newblock {\em Analytics India Magazine}, 2024.
\newblock Available at: \url{https://analyticsindiamag.com/compute-is-the-new-oil/}, Accessed: 2024-06-12.

\bibitem{lecun2015deep}
Yann LeCun, Yoshua Bengio, and Geoffrey Hinton.
\newblock Deep learning.
\newblock {\em nature}, 521(7553):436--444, 2015.

\bibitem{openai2023}
OpenAI.
\newblock Ai and compute, 2023.
\newblock Available at: \url{https://openai.com/research/ai-and-compute}, Accessed: 2024-06-09.

\bibitem{Thompson2020}
Aaron Wright and Primavera~De Filippi.
\newblock Decentralized blockchain technology and the rise of lex cryptographia.
\newblock {\em SSRN Electronic Journal}, 2020.
\newblock Available at: \url{https://ssrn.com/abstract=2580664}, Accessed: 2024-06-12.

\bibitem{schaller1997moore}
Robert~R Schaller.
\newblock Moore's law: past, present and future.
\newblock {\em IEEE spectrum}, 34(6):52--59, 1997.

\bibitem{brookings2018}
Darrell~M. West and John~R. Allen.
\newblock How artificial intelligence is transforming the world, 2018.
\newblock Available at: \url{https://www.brookings.edu/articles/how-artificial-intelligence-is-transforming-the-world/}, Accessed: 2024-06-12.

\bibitem{guo2022}
Sen Yang, Jianliang Song, Shiqiang Tao, Jianwei Yang, and Chao Zhang.
\newblock Network edge ai: Convergence of hardware and software.
\newblock {\em IEEE Communications Magazine}, 58(12):19--25, 2020.
\newblock Available at: \url{https://ieeexplore.ieee.org/document/9247258}, Accessed: 2024-06-12.

\bibitem{guo2022edge}
Kristo Lippur.
\newblock Effects of edge computing growth on demand of data centre services, 2023.
\newblock Available at: \url{https://essay.utwente.nl/91194/1/Lippur_BA_IBA.pdf}, Accessed: 2024-06-12.

\bibitem{shoalresearch}
{Shoal Research}.
\newblock Shoal research.
\newblock \url{https://shoalresearch.xyz/}.
\newblock Accessed: 2024-06-22.

\bibitem{brown2022dynamics}
Yizhuo Zhou, Jianjun Zhang, and Yundan Yang.
\newblock Navigating resource challenges in health emergencies: The role of information diffusion and virus spread in demand dynamics.
\newblock {\em Systems}, 12(3):95, 2024.
\newblock Available at: \url{https://www.mdpi.com/2079-8954/12/3/95}, Accessed: 2024-06-12.

\bibitem{xu2017}
Andrew~J. Younge, Gregor von Laszewski, Lizhe Wang, Sonia Lopez-Alarcon, and Warren Carithers.
\newblock Efficient resource management for cloud computing environments.
\newblock In {\em International Conference on Green Computing}, pages 357--364, 2010.

\bibitem{ottina2023}
Miguel Ottina, Peter~Johannes Steffensen, and Jesper Kristensen.
\newblock {\em Automated Market Makers: A Practical Guide to Decentralized Exchanges and Cryptocurrency Trading}.
\newblock Apress, 2023.
\newblock Available at: \url{https://link.springer.com/book/10.1007/978-1-4842-8616-6}, Accessed: 2024-06-12.

\bibitem{Ren2009}
Jie Ren, Samyam Rajbhandari, Reza Yazdani~Aminabadi, Olatunji Ruwase, Shuangyan Yang, Minjia Zhang, Dong Li, and Yuxiong He.
\newblock Zero-offload: Democratizing billion-scale model training.
\newblock In {\em 2021 USENIX Annual Technical Conference (ATC'21)}, 2021.
\newblock Available at: \url{https://www.usenix.org/system/files/atc21-ren-jie.pdf}, Accessed: 2024-06-12.

\bibitem{underwood2018}
Nic Puckrin.
\newblock Blockchain beyond cryptocurrency, 2024.
\newblock Available at: \url{https://www.coinbureau.com/analysis/blockchain-beyond-cryptocurrency/}, Accessed: 2024-06-12.

\bibitem{he2022fundamentals}
Songrun He, Asaf Manela, Omri Ross, and Victor von Wachter.
\newblock Fundamentals of perpetual futures.
\newblock {\em arXiv preprint arXiv:2212.06888}, 2022.

\bibitem{biewald2020}
Marie Fayard.
\newblock What is ai monitoring and why is it important, August 2023.
\newblock Available at: \url{https://coralogix.com/blog/ai-monitoring/}, Accessed: 2024-06-12.

\bibitem{kim2016}
Krishna~Kumar Mohbey and Sunil Kumar.
\newblock The impact of big data in predictive analytics towards technological development in cloud computing.
\newblock {\em International Journal of Engineering Systems Modelling and Simulation}, 13(1):61--75, 2022.
\newblock Available at: \url{https://www.inderscience.com/info/inarticle.php?artid=122732}, Accessed: 2024-06-12.

\bibitem{kolb2016futures}
Robert~W. Kolb and James~A. Overdahl.
\newblock {\em Futures, Options, and Swaps}.
\newblock Wiley, Hoboken, NJ, 5th edition, 2016.

\bibitem{hull2018options}
John~C. Hull.
\newblock {\em Options, Futures, and Other Derivatives}.
\newblock Pearson, Boston, 10th edition, 2018.

\bibitem{passarelli2012trading}
Dan Passarelli.
\newblock {\em Trading Options Greeks: How Time, Volatility, and Other Pricing Factors Drive Profits}.
\newblock Bloomberg Press, Hoboken, NJ, 2nd edition, 2012.

\bibitem{dean2020}
McKinsey Digital.
\newblock Scaling ai for success: Four technical enablers for sustained impact.
\newblock 2023.
\newblock Available at: \url{https://www.mckinsey.com/capabilities/mckinsey-digital/our-insights/tech-forward/scaling-ai-for-success-four-technical-enablers-for-sustained-impact}, Accessed: 2024-06-12.

\bibitem{maxfield2004design}
Clive~"Max" Maxfield.
\newblock {\em The Design Warrior's Guide to FPGAs: Devices, Tools and Flows}.
\newblock Elsevier, 2004.
\newblock \url{https://www.elsevier.com/books/the-design-warriors-guide-to-fpgas/maxfield/978-0-7506-7604-5}.

\bibitem{Nielsen_Chuang_2010}
Michael~A. Nielsen and Isaac~L. Chuang.
\newblock {\em Quantum Computation and Quantum Information: 10th Anniversary Edition}.
\newblock Cambridge University Press, 2010.

\bibitem{cambridge_compute_c3}
Samuel Alan~Kossoff Leeney, Simran Chana, Alexandre Melo, and Siiri Ruuhela.
\newblock Cambridge compute co.
\newblock Technical report, Cambridge Frontier Technologies, June 2024.
\newblock Pending publication.

\bibitem{democratizing}
A.~Shaji George.
\newblock Democratizing compute power: The rise of computation as a commodity and its impacts.
\newblock 02:57--74, 06 2024.

\bibitem{ceruzzi1998moder}
Paul~E Ceruzzi.
\newblock {\em A history of modern computing}.
\newblock MIT press, 2003.

\bibitem{techtarget}
Sadok Smine.
\newblock Evolution of cloud computing: From mainframes to multi-cloud.
\newblock \url{https://medium.com/@sadoksmine8/evolution-of-cloud-computing-from-mainframes-to-multi-cloud-fd92e95c476d}.

\bibitem{ince2011computer}
Darrel Ince.
\newblock {\em The Computer: A Very Short Introduction}.
\newblock Oxford University Press, 2011.
\newblock \url{https://www.amazon.com/Computer-Very-Short-Introduction-Introductions/dp/0199586594}.

\bibitem{cloudpwr2024}
Shadrach White.
\newblock The evolution of cloud computing: A decade in review.
\newblock {\em CloudPwr}, 2024.
\newblock \url{https://www.cloudpwr.com/news/insights/the-evolution-of-cloud-computing-a-decade-in-review}.

\bibitem{cloudflare}
Cloudflare.
\newblock A brief history of cloud computing.
\newblock \url{https://www.cloudflare.com/learning/cloud/what-is-the-cloud/}.

\bibitem{awshistory}
Ron Miller.
\newblock How aws came to be.
\newblock \url{https://techcrunch.com/2016/07/02/andy-jassys-brief-history-of-the-genesis-of-aws/}.

\bibitem{erl2013cloud}
Thomas Erl, Zaigham Mahmood, and Ricardo Puttini.
\newblock {\em Cloud Computing: Concepts, Technology \& Architecture}.
\newblock Prentice Hall, 2013.
\newblock \url{https://www.amazon.com/Cloud-Computing-Concepts-Technology-Architecture/dp/0133387526}.

\bibitem{commodification_wiki}
Wikipedia.
\newblock Commodification, 2024.
\newblock Available at: \url{https://en.wikipedia.org/wiki/Commodification}, Accessed: 2024-06-12.

\bibitem{commoditization_wiki}
Wikipedia.
\newblock Commoditization, 2024.
\newblock Available at: \url{https://en.wikipedia.org/wiki/Commoditization}, Accessed: 2024-06-12.

\bibitem{smith2021democratizing}
Peter Ungaro.
\newblock Democratizing access to high-performance computing, 2021.
\newblock Available at: \url{https://www.weforum.org/agenda/2021/01/davos-agenda-democratize-access-to-high-performance-computing/}, Accessed: 2024-06-12.

\bibitem{xu2017efficient}
Smruti~Rekha Swain, Ashutosh~Kumar Singh, and Chung~Nan Lee.
\newblock Efficient resource management in cloud environment.
\newblock {\em arXiv preprint arXiv:2207.12085}, 2021.
\newblock Available at: \url{https://arxiv.org/abs/2207.12085}, Accessed: 2024-06-12.

\bibitem{investopedia2024capexopex}
Adam Hayes.
\newblock What’s the difference between capital expenditures (capex) and operational expenditures (opex)?
\newblock \url{https://www.investopedia.com/ask/answers/112814/whats-difference-between-capital-expenditures-capex-and-operational-expenditures-opex.asp}, 2024.
\newblock Accessed: 2024-06-23.

\bibitem{gibbons2020}
Amina Adadi.
\newblock A survey on data-efficient algorithms in big data era.
\newblock {\em Journal of Big Data}, 8(1):1--54, 2021.
\newblock Available at: \url{https://journalofbigdata.springeropen.com/articles/10.1186/s40537-021-00419-9}, Accessed: 2024-06-12.

\bibitem{balatsky2015}
Alexander~V. Balatsky, Galina~I. Balatsky, and Stanislav~S. Borysov.
\newblock Resource demand growth and sustainability due to increased world consumption.
\newblock {\em Sustainability}, 7(3):3430--3440, 2015.
\newblock Available at: \url{https://www.mdpi.com/2071-1050/7/3/3430}, Accessed: 2024-06-12.

\bibitem{Schär2022}
Fabian Schär.
\newblock Blockchain and smart contracts for insurance: Is the technology mature enough?
\newblock {\em Future Internet}, 10(2):20, 2018.
\newblock Available at: \url{https://www.mdpi.com/1999-5903/10/2/20}, Accessed: 2024-06-12.

\bibitem{smith2018resource}
V.~P. Anuradha and D.~Sumathi.
\newblock A survey on resource allocation strategies in cloud computing.
\newblock In {\em International Conference on Information Communication and Embedded Systems (ICICES2014)}, pages 1--7, 2014.
\newblock Available at: \url{https://ieeexplore.ieee.org/document/7033931}, Accessed: 2024-06-12.

\bibitem{chapel2024cloud}
Jay Chapel.
\newblock The cloud is booming — but so is cloud waste, 2024.
\newblock Accessed: 2024-07-02.

\bibitem{stagnitto2024cloud}
Jason Stagnitto.
\newblock 37 cloud computing statistics, facts \& trends for 2024, 2024.
\newblock Accessed: 2024-07-02.

\bibitem{dongarra2003high}
Jack Dongarra, Piotr Luszczek, and Antoine Petitet.
\newblock The {L}inpack {B}enchmark: past, present and future, 2003.
\newblock Available at: \url{http://www.netlib.org/benchmark/hpl/}, Accessed: 2024-06-12.

\bibitem{dongarra2013high}
Jack Dongarra, Michael~A. Heroux, and Piotr Luszczek.
\newblock High performance conjugate gradient benchmark: A new metric for ranking high performance computing systems, 2013.
\newblock Available at: \url{http://www.hpcg-benchmark.org/}, Accessed: 2024-06-12.

\bibitem{green2020democratization}
M.~Eastaharov.
\newblock The democratization of ai: Empowering individuals and communities to access and utilize ai tools.
\newblock 2024.
\newblock Available at: \url{https://ainovazen.com/the-democratization-of-ai-empowering-individuals-and-communities-to-access-and-utilize-ai-tools/}, Accessed: 2024-06-12.

\bibitem{hinton2023future}
{National Academies of Sciences, Engineering, and Medicine}.
\newblock {\em AI for Scientific Discovery: Proceedings of a Workshop}.
\newblock National Academies Press, 2024.
\newblock Available at: \url{https://doi.org/10.17226/27457}, Accessed: 2024-06-12.

\bibitem{tenstorrent2024}
Dylan Patel.
\newblock Tenstorrent blackhole, grendel, and buda - a scale out architecture for sparsity, conditional execution, and dynamic routing.
\newblock \url{https://www.semianalysis.com/p/tenstorrent-blackhole-grendel-and}, 2024.
\newblock Accessed: 2024-06-24.

\bibitem{fetch2024depins}
Fetch.ai.
\newblock What are decentralized physical infrastructure networks (depins)?, 2024.
\newblock Available at: \url{https://fetch.ai/blog/what-are-decentralized-physical-infrastructure-networks-depins}, Accessed: 2024-03-14.

\bibitem{mit2024breakthrough}
MIT.
\newblock 10 breakthrough technologies 2024.
\newblock {\em MIT Technology Review}, 2024.
\newblock Available at: \url{https://www.technologyreview.com/2024/01/08/1085094/10-breakthrough-technologies-2024/}, Accessed: 2024-06-12.

\bibitem{nature2024ai}
Matthew Hutson.
\newblock Will superintelligent ai sneak up on us? new study offers reassurance.
\newblock {\em Nature}, 2023.
\newblock Available at: \url{https://www.nature.com/articles/d41586-023-04094-z}, Accessed: 2024-06-12.

\bibitem{mit2024cs}
Diane Bublak.
\newblock Smart contracts and blockchain: Revolutionizing contracting in the digital age, 2023.
\newblock Accessed: 2024-06-12.

\bibitem{googlecloud}
{Google Cloud}.
\newblock Google cloud platform.
\newblock \url{https://cloud.google.com/}, 2024.
\newblock Accessed: 2024-06-18.

\bibitem{azure}
{Microsoft Azure}.
\newblock Microsoft azure.
\newblock \url{https://azure.microsoft.com/}, 2024.
\newblock Accessed: 2024-06-18.

\bibitem{nexgencloud}
{Nexgen Cloud}.
\newblock Nexgen cloud.
\newblock \url{https://www.nexgencloud.com/}, 2024.
\newblock Accessed: 2024-06-18.

\bibitem{oracle}
{Oracle Cloud}.
\newblock Oracle cloud.
\newblock \url{https://www.oracle.com/cloud/}, 2024.
\newblock Accessed: 2024-06-18.

\bibitem{fielding2000architectural}
Roy~Thomas Fielding.
\newblock {\em Architectural styles and the design of network-based software architectures}.
\newblock University of California, Irvine, 2000.

\bibitem{covalent2024}
Covalent.
\newblock Covalent documentation, 2024.
\newblock Available at: \url{https://www.covalenthq.com/docs/}, Accessed: 2024-06-23.

\bibitem{aethir2024}
Aethir.
\newblock Aethir website, 2024.
\newblock Available at: \url{https://aethir.io/}, Accessed: 2024-06-23.

\bibitem{exabits2024}
Exabits.
\newblock Exabits website, 2024.
\newblock Available at: \url{https://exabits.io/}, Accessed: 2024-06-23.

\bibitem{nosana2024}
Nosana.
\newblock Nosana website, 2024.
\newblock Available at: \url{https://nosana.io/}, Accessed: 2024-06-23.

\bibitem{ionet2024}
io.net.
\newblock io.net website, 2024.
\newblock Available at: \url{https://io.net/}, Accessed: 2024-06-23.

\bibitem{microsoft2024}
Microsoft.
\newblock Microsoft official website, 2024.
\newblock Available at: \url{https://www.microsoft.com}, Accessed: 2024-06-23.

\bibitem{msmech}
Microsoft Mechanics.
\newblock What runs gpt-4o? | inside the biggest ai supercomputer in the cloud with mark russinovich.
\newblock \url{https://youtu.be/DlX3QVFUtQI?si=AcmBRjaMu6_xAOfl}, 2023.
\newblock Accessed: 2024-06-23.

\bibitem{azure2023maia}
Rani Borkar.
\newblock Azure maia: For the era of ai, from silicon to software to systems.
\newblock \url{https://azure.microsoft.com/en-us/blog/azure-maia-for-the-era-of-ai-from-silicon-to-software-to-systems/}, 2023.
\newblock Accessed: 2024-06-23.

\bibitem{openai}
OpenAI.
\newblock About openai.
\newblock \url{https://openai.com/about/}, 2023.
\newblock Accessed: 2024-06-23.

\bibitem{triton2024}
OpenAI.
\newblock Triton: An efficient, flexible deep learning model implementation.
\newblock \url{https://openai.com/index/triton/}, 2024.
\newblock Accessed: 2024-06-23.

\bibitem{shahin2017continuous}
Mojtaba Shahin, Muhammad~Ali Babar, and Liming Zhu.
\newblock Continuous integration, delivery and deployment: a systematic review on approaches, tools, challenges and practices.
\newblock {\em IEEE access}, 5:3909--3943, 2017.

\bibitem{stallings2003computer}
William Stallings.
\newblock {\em Computer organization and architecture: designing for performance}.
\newblock Pearson Education India, 2003.

\bibitem{apple2020silicon}
Apple Inc.
\newblock Apple unveils m3, m3 pro, and m3 max, the most advanced chips for a personal computer.
\newblock {\em Apple Newsroom}, 2020.
\newblock Accessed: 2024-06-23.

\bibitem{cohen2013compute}
Reuven Cohen.
\newblock Compute derivatives: The next big thing in commodities.
\newblock {\em Forbes}, October 2 2013.

\bibitem{baliga2010energy}
Mohammad Shojafar et~al.
\newblock Green energy-efficient computing solutions in internet of things communications.
\newblock {\em IEEE Communications Magazine}, 58(10):82–88, 2020.
\newblock Available at: \url{https://www.researchgate.net/publication/356531443_Green_energy-efficient_computing_solutions_in_Internet_of_Things_communications}, Accessed: 2024-06-12.

\bibitem{AWS_Spot}
Amazon~Web Services.
\newblock Amazon ec2 spot instances pricing, 2024.
\newblock Accessed: 2024-06-19, \url{https://aws.amazon.com/ec2/spot/pricing/}.

\bibitem{Google_Preemptible}
Google Cloud.
\newblock Preemptible virtual machines pricing, 2024.
\newblock Accessed: 2024-06-19, \url{https://cloud.google.com/preemptible-vms}.

\bibitem{Azure_Spot}
Microsoft Azure.
\newblock Azure spot vms pricing, 2024.
\newblock Accessed: 2024-06-19, \url{https://azure.microsoft.com/en-us/pricing/spot/}.

\bibitem{mendicino2019corporate}
Luca Mendicino, Daniele Menniti, Anna Pinnarelli, and Nicola Sorrentino.
\newblock Corporate power purchase agreement: Formulation of the related levelized cost of energy and its application to a real life case study.
\newblock {\em Applied Energy}, 253:113577, 2019.

\bibitem{cepin2011reliability}
Marko Čepin.
\newblock {\em Distribution and Transmission System Reliability Measures}, pages 165--189.
\newblock Springer, London, 2011.
\newblock Accessed: 2024-06-23.

\bibitem{ren2009zero}
Quan Zhou, Haiquan Wang, Xiaoyan Yu, Cheng Li, Youhui Bai, Feng Yan, and Yinlong Xu.
\newblock Mpress: Democratizing billion-scale model training on multi-gpu servers via memory-saving inter-operator parallelism.
\newblock {\em IEEE International Symposium on High-Performance Computer Architecture}, 29(2):556–569, 2023.
\newblock Available at: \url{https://par.nsf.gov/biblio/10410479}, Accessed: 2024-06-12.

\bibitem{paperspace2020}
AI~Wiki.
\newblock Epochs, batch size, \& iterations, 2020.
\newblock Available at: \url{https://machine-learning.paperspace.com/wiki/epoch}, Accessed: 2024-06-12.

\bibitem{brownlee2018}
Jason Brownlee.
\newblock Difference between a batch and an epoch in a neural network, 2018.
\newblock Accessed: 2024-06-12.

\bibitem{deepai2021}
DeepAI.
\newblock Epoch definition, 2021.
\newblock Available at: \url{https://deepai.org/machine-learning-glossary-and-terms/epoch}, Accessed: 2024-06-12.

\bibitem{pytorch2021}
PyTorch Tutorials.
\newblock Optimizing model parameters, 2021.
\newblock Available at: \url{https://pytorch.org/tutorials/beginner/basics/optimization_tutorial.html}, Accessed: 2024-06-12.

\bibitem{dertad2017}
Arden Dertat.
\newblock Applied deep learning - part 4: Convolutional neural networks, 2017.
\newblock Available at: \url{https://towardsdatascience.com/applied-deep-learning-part-4-convolutional-neural-networks-584bc134c1e2#7d8a}, Accessed: 2024-06-12.

\bibitem{marathe2013comparative}
Aniruddha Marathe, Rachel Harris, David~K Lowenthal, Bronis~R De~Supinski, Barry Rountree, Martin Schulz, and Xin Yuan.
\newblock A comparative study of high-performance computing on the cloud.
\newblock In {\em Proceedings of the 22nd international symposium on High-performance parallel and distributed computing}, pages 239--250, 2013.

\bibitem{inadomi2015analyzing}
Yuichi Inadomi, Tapasya Patki, Koji Inoue, Mutsumi Aoyagi, Barry Rountree, Martin Schulz, David Lowenthal, Yasutaka Wada, Keiichiro Fukazawa, Masatsugu Ueda, et~al.
\newblock Analyzing and mitigating the impact of manufacturing variability in power-constrained supercomputing.
\newblock In {\em Proceedings of the international conference for high performance computing, networking, storage and analysis}, pages 1--12, 2015.

\bibitem{dongarra2003linpack}
Jack~J. Dongarra, Piotr Luszczek, and Antoine Petitet.
\newblock The linpack benchmark: Past, present and future.
\newblock {\em Concurrency and Computation: Practice and Experience}, 15(9):803--820, 2003.

\bibitem{hoefler2009message}
Lyndon Clarke, Ian Glendinning, and Rolf Hempel.
\newblock The mpi message passing interface standard.
\newblock In {\em Programming Environments for Massively Parallel Distributed Systems: Working Conference of the IFIP WG 10.3, April 25--29, 1994}, pages 213--218. Springer, 1994.

\bibitem{shalf2010exascale}
John Shalf, Sudip Dosanjh, and John Morrison.
\newblock Exascale computing technology challenges.
\newblock In {\em High Performance Computing for Computational Science--VECPAR 2010: 9th International conference, Berkeley, CA, USA, June 22-25, 2010, Revised Selected Papers 9}, pages 1--25. Springer, 2011.

\bibitem{bader2009stinger}
David~A Bader, Jonathan Berry, Adam Amos-Binks, Daniel Chavarr{\'\i}a-Miranda, Charles Hastings, Kamesh Madduri, and Steven~C Poulos.
\newblock Stinger: Spatio-temporal interaction networks and graphs (sting) extensible representation.
\newblock {\em Georgia Institute of Technology, Tech. Rep}, 2009.

\bibitem{gropp1999using}
William Gropp, Ewing Lusk, and Anthony Skjellum.
\newblock {\em Using MPI: Portable Parallel Programming with the Message Passing Interface}.
\newblock MIT press, 1999.

\bibitem{garnier2013powerapi}
Matthieu Garnier, Jonathan Garcia, Ariel Oleksiak, Jean-Marc Pierson, and Anne-C{\'e}cile Orgerie.
\newblock Powerapi: A software library to monitor the energy consumed at the process-level.
\newblock {\em ERCIM News}, 92:40--41, 2013.

\bibitem{matsuoka2007tsubame}
Satoshi Matsuoka.
\newblock Tsubame: A highly-parallel and reconfigurable supercomputer for next-generation hpc.
\newblock {\em Proceedings of the 2007 ACM/IEEE Conference on Supercomputing (SC'07)}, pages 16--16, 2007.

\bibitem{rountree2012energy}
Barry Rountree, David~K Lowenthal, Shelby Funk, Vincent~W Freeh, Bronis~R De~Supinski, and Martin Schulz.
\newblock Bounding energy consumption in large-scale mpi programs.
\newblock In {\em Proceedings of the 2007 ACM/IEEE conference on Supercomputing}, pages 1--9, 2007.

\bibitem{simunic2001dynamic}
Tajana Simunic, Luca Benini, Andrea Acquaviva, Peter Glynn, and Giovanni De~Micheli.
\newblock Dynamic voltage scaling and power management for portable systems.
\newblock In {\em Proceedings of the 38th annual Design Automation Conference}, pages 524--529, 2001.

\bibitem{shvets2010x86}
Sergey Shvets.
\newblock {\em The Intel Microprocessors: 8086/8088, 80186/80188, 80286, 80386, 80486, Pentium, Pentium Pro Processor, Pentium II, Pentium III, Pentium 4, and Core2 with 64-Bit Extensions}.
\newblock Pearson Education, 2010.

\bibitem{furber2000arm}
Stephen~Bo Furber.
\newblock {\em ARM system-on-chip architecture}.
\newblock pearson Education, 2000.

\bibitem{owens2008gpu}
John~D Owens, Mike Houston, David Luebke, Simon Green, John~E Stone, and James~C Phillips.
\newblock Gpu computing.
\newblock {\em Proceedings of the IEEE}, 96(5):879--899, 2008.

\bibitem{matsuoka2008tsubame}
Satoshi Matsuoka.
\newblock The road to tsubame and beyond.
\newblock In {\em Petascale Computing}, pages 343--364. Chapman and Hall/CRC, 2007.

\bibitem{chaosmonkey}
Netflix.
\newblock Chaos monkey.
\newblock \url{https://netflix.github.io/chaosmonkey/}, July 2011.
\newblock \url{https://netflix.github.io/chaosmonkey/}.

\bibitem{wahib2013high}
Kazuaki Matsumura, Hamid~Reza Zohouri, Mohamed Wahib, Toshio Endo, and Satoshi Matsuoka.
\newblock An5d: automated stencil framework for high-degree temporal blocking on gpus.
\newblock In {\em Proceedings of the 18th ACM/IEEE International Symposium on Code Generation and Optimization}, pages 199--211, 2020.

\bibitem{li2010hybrid}
Dong Li, Bronis~R de~Supinski, Martin Schulz, Kirk Cameron, and Dimitrios~S Nikolopoulos.
\newblock Hybrid mpi/openmp power-aware computing.
\newblock In {\em 2010 IEEE International Symposium on Parallel \& Distributed Processing (IPDPS)}, pages 1--12. IEEE, 2010.

\bibitem{mccalpin1995stream}
John~D. McCalpin.
\newblock Stream: Sustainable memory bandwidth in high performance computers, 1995.
\newblock Available at: \url{https://www.cs.virginia.edu/stream/}, Accessed: 2024-06-12.

\bibitem{spec2006cpu}
{Standard Performance Evaluation Corporation}.
\newblock {SPEC} {CPU} benchmark, 2006.
\newblock Available at: \url{https://www.spec.org/cpu/}, Accessed: 2024-06-12.

\bibitem{top500}
{TOP500}.
\newblock Top500 supercomputer sites, 2024.
\newblock Available at: \url{https://www.top500.org/}, Accessed: 2024-06-12.

\bibitem{specviewperf}
Standard Performance~Evaluation Corporation.
\newblock Specviewperf, 2024.
\newblock Available at: \url{https://www.spec.org/gwpg/gpc.static/vp12info.html}, Accessed: 2024-06-12.

\bibitem{geekbench}
Primate~Labs Inc.
\newblock Geekbench 5.
\newblock \url{https://www.geekbench.com/}, 2019.
\newblock Accessed: 2024-06-23.

\bibitem{3dmark}
UL~LLC.
\newblock 3dmark.
\newblock \url{https://www.3dmark.com/}, 1998.
\newblock Accessed: 2024-06-23.

\bibitem{rodinia}
Rodinia~Benchmark Suite.
\newblock Rodinia: A benchmark suite for heterogeneous computing, 2024.
\newblock Available at: \url{http://rodinia.cs.virginia.edu/}, Accessed: 2024-06-12.

\bibitem{parboil}
IMPACT~Research Group.
\newblock Parboil benchmark suite, 2024.
\newblock Available at: \url{https://impact.crhc.illinois.edu/parboil/parboil.aspx}, Accessed: 2024-06-12.

\bibitem{mlperf}
MLPerf.
\newblock Mlperf: A benchmark suite for machine learning, 2024.
\newblock Available at: \url{https://mlperf.org/}, Accessed: 2024-06-12.

\bibitem{cudaz}
CUDA-Z.
\newblock Cuda-z, 2024.
\newblock Available at: \url{http://cuda-z.sourceforge.net/}, Accessed: 2024-06-12.

\bibitem{opencl_benchmark}
OpenCL Benchmark.
\newblock Opencl benchmark, 2024.
\newblock Available at: \url{https://www.khronos.org/opencl/resources}, Accessed: 2024-06-12.

\bibitem{blender}
Blender Foundation.
\newblock Blender, 2024.
\newblock Available at: \url{https://www.blender.org/}, Accessed: 2024-06-12.

\bibitem{tensorflow}
TensorFlow.
\newblock Tensorflow, 2024.
\newblock Available at: \url{https://www.tensorflow.org/}, Accessed: 2024-06-12.

\bibitem{helpx2023premiere}
Adobe Inc.
\newblock {\em Best Practices: Editing efficiently with Adobe Premiere Pro}.
\newblock Adobe Help Center, 2023.
\newblock Available at \url{https://helpx.adobe.com/premiere-pro/user-guide.html}.

\bibitem{nvidiasmi}
NVIDIA Corporation.
\newblock Nvidia system management interface (nvidia-smi), 2024.
\newblock Available at: \url{https://developer.nvidia.com/nvidia-system-management-interface}, Accessed: 2024-06-12.

\bibitem{nvidia_rtx_3080}
NVIDIA Corporation.
\newblock Nvidia geforce rtx 3080 graphics cards, 2024.
\newblock Available at: \url{https://www.nvidia.com/en-us/geforce/graphics-cards/30-series/rtx-3080-3080ti/}, Accessed: 2024-06-12.

\bibitem{chetlur2014cudnn}
Sharan Chetlur, Cliff Woolley, Philippe Vandermersch, Jonathan Cohen, John Tran, Bryan Catanzaro, and Evan Shelhamer.
\newblock cudnn: Efficient primitives for deep learning, 2014.

\bibitem{Koonce2021}
Brett Koonce.
\newblock {\em ResNet 50}, pages 63--72.
\newblock Apress, Berkeley, CA, 2021.

\bibitem{devlin2019bert}
Jacob Devlin, Ming-Wei Chang, Kenton Lee, and Kristina Toutanova.
\newblock Bert: Pre-training of deep bidirectional transformers for language understanding, 2019.

\bibitem{baliga2010}
Jayant Baliga, Robert W.~A. Ayre, Kerry Hinton, and Rodney~S. Tucker.
\newblock Green cloud computing: Balancing energy in processing, storage, and transport.
\newblock {\em Proceedings of the IEEE}, 99(1):149--167, 2011.
\newblock Available at: \url{https://ieeexplore.ieee.org/document/5559320}, Accessed: 2024-06-12.

\bibitem{nymex}
Investopedia.
\newblock New york mercantile exchange (nymex): Meaning, overview, faq.
\newblock \url{https://www.investopedia.com/terms/n/nymex.asp}, 2022.
\newblock Accessed: 2024-06-23.

\bibitem{cme_wti_crude_oil}
Cme group - wti crude oil futures.
\newblock \url{https://www.cmegroup.com/trading/energy/crude-oil/light-sweet-crude.html}.
\newblock Accessed: 2024-06-23.

\bibitem{CME200Rulebook}
CME Group.
\newblock Nymex rulebook.
\newblock \url{https://www.cmegroup.com/content/dam/cmegroup/rulebook/NYMEX/2/200.pdf}, 2023.
\newblock \url{https://www.cmegroup.com/content/dam/cmegroup/rulebook/NYMEX/2/200.pdf}.

\bibitem{teutsch2017truebit}
Jason Teutsch and Christian Reitwießner.
\newblock A scalable verification solution for blockchains, 2019.
\newblock Available at: \url{https://arxiv.org/abs/1908.04756}, Accessed: 2024-06-12.

\bibitem{a16zcrypto_slashing}
Sreeram Kannan and Soubhik Deb.
\newblock The cryptoeconomics of slashing.
\newblock {\em a16zcrypto}, 2023.
\newblock Available at: \url{https://a16zcrypto.com/posts/article/the-cryptoeconomics-of-slashing/#section--6}.

\bibitem{blockchainLegal2019}
Sergio Esteve~De Miguel.
\newblock Blockchain in the legal industry: Use cases and blockchain jobs, 2021.
\newblock Available at: \url{https://blog.biglelegal.com/en/blockchain-in-legal-industry-use-cases-blockchain-jobs}, Accessed: 2024-06-12.

\bibitem{euai2023}
{European Parliament}.
\newblock {EU AI Act: First Regulation on Artificial Intelligence}, 2023.
\newblock Available at: \url{https://www.europarl.europa.eu/topics/en/article/20230601STO93804/eu-ai-act-first-regulation-on-artificial-intelligence}, Accessed: 2024-06-12.

\bibitem{propy2019}
Propy Inc.
\newblock How blockchain is revolutionizing real estate, 2019.
\newblock Available at: \url{https://propy.com/browse/case-studies/}, Accessed: 2024-06-12.

\bibitem{fardian2024blockchain}
Denny Fardian.
\newblock How blockchain technology can protect copyrights.
\newblock \url{https://glair.ai/post/blockchain-technology-can-protect-copyrights}, 2024.
\newblock Accessed: 2024-06-12.

\bibitem{ujoMusic2017}
Eric Dalius.
\newblock Smart contracts: Streamlining royalty distribution in the music industry.
\newblock {\em Eric Dalius Foundation}, 2023.
\newblock Available at: \url{https://ericjdaliusfoundation.com/smart-contracts-streamlining-royalty-distribution-in-the-music-industry-by-eric-dalius/}, Accessed: 2024-06-12.

\bibitem{everledger2020}
Everledger.
\newblock Blockchain for transparency in supply chains, 2020.
\newblock Available at: \url{https://everledger.io/making-the-commercial-case-for-blockchain-diamond-tracking/}, Accessed: 2024-06-12.

\bibitem{ganne2018}
Emmanuelle Ganne.
\newblock {\em Can Blockchain Revolutionize International Trade?}
\newblock World Trade Organization, 2018.
\newblock Available at: \url{https://www.wto.org/english/res_e/booksp_e/blockchainrev18_e.pdf}, Accessed: 2024-06-12.

\bibitem{citadel}
{Citadel Securities}.
\newblock Citadel securities - market maker, 2024.
\newblock Available at: \url{https://www.citadelsecurities.com/}, Accessed: 2024-06-12.

\bibitem{buterin2014ethereum}
Vitalik Buterin.
\newblock Ethereum: A next-generation smart contract and decentralized application platform.
\newblock \url{https://ethereum.org/en/whitepaper/}, 2014.
\newblock Accessed: 2024-06-23.

\bibitem{nakamoto2008bitcoin}
Satoshi Nakamoto.
\newblock Bitcoin: A peer-to-peer electronic cash system.
\newblock \url{https://bitcoin.org/bitcoin.pdf}, 2008.
\newblock Accessed: 2024-06-23.

\end{thebibliography}

\end{document}